\documentclass[11pt]{article}%
\usepackage{graphics,graphicx}
\usepackage{amsmath,amssymb,amsfonts,mathrsfs}
\usepackage{color}
\usepackage{epsfig}%
\usepackage{hyperref}

\newtheorem{theorem}{Theorem}

\newtheorem{definition}{Definition}

\newtheorem{remark}{Remark}

\usepackage{makeidx}

\begin{document}

\title{Quantum Feedback Networks and  Control: A Brief Survey
\thanks{This work was partially supported by the National Natural Science Foundation of China under Grant No. 60804015, RGC PolyU 5203/10E, and AFOSR Grant FA2386-09-1-4089 AOARD 094089.}}

\author{Guofeng~Zhang\thanks{%
G. Zhang was with the Research School of of Engineering, Australian National University, Canberra, ACT 0200, Australia. He is now with the Department of Applied Mathematics, Hong Kong Polytechnic University, Hong Kong, China (e-mail: magzhang@polyu.edu.hk). } ~~ Matthew~R.~James\thanks{M. R. James is with the ARC Centre for Quantum Computation and Communication Technology, Research School of Engineering, Australian National University, Canberra, ACT 0200, Australia (e-mail: Matthew.James@anu.edu.au).}
}
\date{}
\maketitle

\begin{abstract}
The purpose of this paper is to provide a brief review of some recent developments in quantum  feedback networks and control. A quantum feedback network (QFN) is an interconnected system consisting of open quantum systems linked by free fields and/or direct physical couplings. Basic network constructs, including series connections as well as feedback loops, are discussed. The quantum feedback network theory provides a natural framework for analysis and design. Basic   properties such as dissipation, stability, passivity and gain of open quantum systems are discussed. Control system design is also discussed, primarily in the context of open linear quantum stochastic systems. The issue of physical realizability is discussed, and explicit criteria for stability, positive real lemma, and bounded real lemma are presented. Finally for linear quantum systems, coherent $H^\infty$ and LQG control are described.

\end{abstract}

\tableofcontents

\section{Introduction}

Quantum technology is an interdisciplinary field that studies how to engineer devices
by exploiting their quantum features. Regarded as the second quantum
revolution, quantum technology has many potential far-reaching
applications \cite{DM03}. For example, Shor \cite{Sho94} presented a
quantum algorithm which can offer exponential speedup over
classical algorithms for factoring large integers into prime
numbers. Bennett et al. \cite{BBC+93} proposed a quantum
teleportation protocol where an unknown quantum state can be
disembodiedly transported to a desired receiver. Atomic lasers hold
promising applications in nanotechnology such as atom
lithography, atom optics and precision measurement \cite{MAK+97}.
Quantum technology (including quantum information technology) has
more powerful capability than traditional technology and is one of
the main focuses of scientists. Nevertheless, many challenging problems require to
be systematically presented and successfully addressed in order to foster wider real-world
applications of quantum technology in our life \cite{DM03}.

Recent years have seen a rapid growth of quantum feedback control theory  \cite{MK05,BvHJ07,DP10,BCR10,JK10,WM10}. If measurement is involved in the feedback loop, the feedback mechanism is conventionally called \textit{measurement-based
feedback}, e.g., \cite{Belavkin83,WM93,DJ99,DHJMT00,AASDM02,vHSM05,MvH07,WM10}. Measurement-based feedback control of quantum systems is important in  a
number of areas of quantum technology, including quantum optical
systems, nano-mechanical systems, and circuit QED systems. In measurement-based feedback
control, the plant is a quantum system, while the controller is a
classical (namely non-quantum) system. The classical controller
processes the outcomes of measurement of an observable of the
quantum system (e.g. the number of photons of an optical field) to
determine the classical control actions (e.g. magnetic field) that are applied to control
the behavior of the quantum system. Classical controllers are typically implemented using standard
analog or digital electronics. However, for quantum systems
that have bandwidth much higher than that of conventional
electronics, an important practical issue for the
implementation of measurement-based feedback control systems is the
relatively slow speed of standard classical electronics, since the
feedback system will not work properly unless the controller is
fast enough.

Alternatively, quantum components may be connected to
each other without any measurement devices in the interconnections.
For example, two optical cavities can be connected via
electromagnetic fields (light beams). Such feedback mechanism is referred to as
\textit{coherent feedback} as originally proposed in \cite{YD84,WM94b,NWC+00,Lloyd00,YK03a,YK03b}. The interconnection of a quantum plant and a quantum controller produces a fully quantum system; quantum information flows in this coherent feedback network, thus coherence is preserved in the whole quantum network. Moreover, a coherent feedback controller may have the similar time scale as the plant, and likely would be much faster than classical signal processing. Finally, it is becoming feasible to implement quantum networks in semiconductor materials, for example, photonic crystals are periodic optical nanostructures that are designed to affect the motion of photons in a similar way that periodicity of a semiconductor crystal affects the motion of electrons, and it may be desirable to implement control networks on the same chip (rather than interfacing to a separate system), \cite{MPS+09,PMO+09,SKT+09,SSV+10,TNP+11}. For several reasons, then, it is desirable to implement controllers using the same or similar (e.g. in time scales) hardware as plants. Therefore, it might be advantageous to design coherent feedback networks.

Recently there is a growing interest in the study of coherent quantum feedback networks and control. For example, quantum feedback network structure has been studied \cite{WM94b,YK03a,GJ09,GJ09b,JG10,GJN10,ZJ11}. The problem of $H^\infty$ control has been discussed \cite{JNP08,Mabuchi08,MP11,ZJ11}. The problem of coherent LQG control has been investigated\cite{NJP09,ZJ11}. The issue of physical realizability has been analyzed \cite{JNP08,SP09,MP11a,ZJ11}. The problem of network synthesis of quantum systems via optical devices have been studied \cite{NJD09,Nur10a,Nur10b,petersen11}.  There are also many papers investigating the applications of coherent feedback control, such as intra-cavity squeezing \cite{WM94b}, optical field squeezing \cite{YK03b,GW09,IYY+11}, $H^\infty$ control \cite{Mabuchi08}, entanglement enhancement \cite{HP10}, polarization squeezing \cite{SM06,SSM08}, error correction in quantum memories \cite{KNP+10,KPC+11}, and optical switches \cite{Mabuchi11}.

The paper is organized as follows. Section \ref{sec:models-closed} discusses closed quantum systems, in particular, closed quantum harmonic oscillators. Section \ref{sec:field_systems} introduces Boson field and a basic model structure of open quantum systems. Section \ref{sec:interconnection} discusses mechanisms by which open quantum systems interconnect. Section \ref{sec:models-general} focuses on linear quantum systems. Section \ref{sec:general_sytemes} presents results for quantum dissipative systems. Section \ref{sec:performance_specification} discusses fundamental characteristics of linear quantum systems. Section \ref{sec:feedback} presents $H^\infty$ and LQG controller synthesis of linear quantum systems. Section \ref{sec:synthesis} touches on how linear quantum systems can be realized by means of optical devices. Section \ref{sec:conclusion} concludes the paper.

\textbf{Notation}. $i$ is the imaginary unit. $\delta_{jk}$ is Kronecker delta, and $\delta(t)$ is Dirac delta. Given a column vector of operators or complex numbers $x=[
\begin{array}{ccc}
x_{1} & \cdots & x_{m}%
\end{array}%
]^{T}$ where $m$ is a positive integer, define $x^{\#}=[
\begin{array}{ccc}
x_{1}^{\ast } & \cdots & x_{m}^{\ast }%
\end{array}%
]^{T}$, where the asterisk $\ast $ indicates Hilbert space adjoint or complex conjugation. Denote $x^\dag = (x^\#)^T =[
\begin{array}{ccc}
x_{1}^{\ast } & \cdots & x_{m}^{\ast }%
\end{array}%
]$. Furthermore, define the doubled-up column vector to be $%
\breve{x}=[
\begin{array}{cc}
x^{T} & \left( x^{\#}\right) ^{T}%
\end{array}%
]^{T}$. The matrix case can be defined analogously. Given two matrices $U$, $V\in \mathbb{C}^{r\times k}$, a doubled-up matrix $\Delta \left( U,V\right) $ is defined as $\Delta \left( U,V\right) :=[
\begin{array}{cccc}
U & V; & V^{\#} & U^{\#}%
\end{array}%
] $. Let $I_{n}$ be an identity matrix. Define $J_{n}={\rm diag}(I_{n},-I_{n})$ and $\Theta_{n}=[0 ~ ~I_{n}; ~ -I_{n} ~~ 0 ]$. (The subscript ``$n$'' is always omitted.) Then for a matrix $X\in \mathbb{C}^{2n\times 2m}$, define $X^{\flat }:=J_{m}X^{\dagger }J_{n}$. Finally we also use $I$ to denote identity operators.

\section{Closed Systems} \label{sec:models-closed}
In this paper, closed systems means systems that have no interactions with other systems and/or environment. In this section starting from the fundamental Schrodinger's equation for closed quantum systems, we introduce closed quantum harmonic oscillators which in later sections will be allowed to interact with other systems or electromagnetic fields to produce open quantum systems.

Given a closed quantum system with Hamiltonian $H$, we have the following Schrodinger's equation \footnote{The reduced Planck constant $\bar{h}$ is omitted throughout the paper.}
\begin{equation}\label{eq:U}
    \frac{d}{dt}U(t) = -iHU(t), ~~~ U(0)=I.
\end{equation}
Clearly, $U(t)$ is a unitary operator. The system variables $X(t)$ evolve according to $X(t) = U^\ast(t)XU(t)$ with initial point $X(0)=X$, which satisfy, the Heisenberg picture,
\begin{equation}\label{eq:X_Heisenberg}
 \frac{d}{dt}X(t) = -i [X(t), H(t)].
\end{equation}
(Note that for closed systems $H(t)\equiv H$ for all $t$ due to preservation of energy).

Alternatively, the system density operator $\rho(t) = U(t)\rho U^\ast(t)$ with $\rho(0) = \rho$ satisfies, the Schrodinger picture,
\begin{equation}\label{eq:rho}
 \frac{d}{dt}\rho(t) = -i [H, \rho(t)].
\end{equation}

\subsection{Closed Quantum Harmonic Oscillators}\label{sec:closed_qho}

\begin{figure}[tbh]
\centering
  \centering
  \includegraphics[width=1.5in]{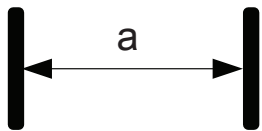}\\
  \caption{Closed optical cavity, black rectangles denote fully reflecting mirrors at cavity resonant frequency}\label{fig_closed_cavity}
\end{figure}

An example of closed quantum harmonic oscillators is an optical cavity with $H = \omega a^\ast a$ (upon scaling), Fig. \ref{fig_closed_cavity}, where $\omega$ is the resonant frequency, and the annihilation operator $a$ is the cavity mode (an operator on a Hilbert space). The adjoint operator $a^\ast$ of $a$ is called the creation operator. $a$ and $a^\ast$ satisfy the canonical commutation relation $[a(t), a^\ast(t)] = 1$ for all $t \geq 0$.  Finally by Eq. (\ref{eq:X_Heisenberg}),
\begin{equation}\label{eq:X2}
 \frac{d}{dt}a(t) = -i\omega a(t), ~~ a(0) = a.
\end{equation}
Therefore $a(t) = e^{-i\omega t}a$ --- an oscillator.

In general, let $G$ be a closed quantum system of interconnection of $n$ quantum harmonic oscillators. The behavior of $G$  is determined by the Hamiltonian
\begin{equation} \label{eq:H0}
H =\frac{1}{2}\breve{a}^{\dagger }\left[
\begin{array}{cc}
\Omega _{-} & \Omega _{+} \\
\Omega _{+}^{\#} & \Omega _{-}^{\#}%
\end{array}%
\right] \breve{a},
\end{equation}%
where $\Omega _{-}$ and $\Omega _{+}$ are respectively $\mathbb{C}^{n  \times n }$ matrices satisfying $\Omega _{-}=\Omega_{-}^{\dagger }$ and $\Omega _{+}=\Omega _{+}^{T}$. By Eq. (\ref{eq:X_Heisenberg}),
\[
\dot a_j (t) =-i [a_j(t) , H(t) ], ~ a_j(0)=a_j, ~~ (j=1,\ldots,n).
\]
In a compact form we have the following linear differential equations
\begin{equation}
\dot{\breve{a}}(t)  = A_0 \breve{a}(t)
\label{eq:closed-3}
\end{equation}
with initial condition $\breve{a}(0)=\breve{a}$, where
\begin{equation}
A_0 = - \Delta( i\Omega_-, i \Omega_+).
\label{eq:A-1}
\end{equation}

\section{Quantum Fields and Open Quantum Systems} \label{sec:field_systems}

\subsection{Boson Fields} \label{sec:field}
The $m$-channel Boson field $b(t) = [b_1(t), \ldots, b_m(t)]^T$ are operators on a Fock space $\mathcal{F}$ \cite{Pa92}, whose components satisfy the singular commutation relations
\begin{equation}\label{eq:CCR}
[ b_j  (t) , b_k^\ast(t') ] = \delta_{jk} \delta(t-t'), ~ [ b_j  (t) , b_k(t') ]= 0, ~ [ b_j^\ast(t) , b_k^\ast(t') ]=0 , ~ (j,k=1, \ldots, m) .
\end{equation}
The operators $b_j(t)$ may be regarded as quantum stochastic processes, see, eg., \cite[Chapter 5]{GZ04}; when the field is in the vacuum state, namely absolutely zero temperature and completely dark, they are called standard \emph{quantum white noise} (that is, $M=N=0$ in \cite[Eq. (10.2.38)]{GZ04}). The integrated processes $B_j(t) = \int_0^t b_j(\tau) d\tau$ are \emph{quantum Wiener processes} with Ito increments $dB_j(t) = B_j(t+dt)-B_j(t)$, $(j=1,\ldots,m)$.

There might exist scattering between channels, which is modeled by the gauge process
\begin{equation} \label{gauge}
\Lambda(t) = \int_0^t b^\#(\tau)b^T(\tau)d\tau =  \left[ \begin{array}{ccc}
                                  \Lambda_{11}(t) & \cdots  & \Lambda_{1m}(t) \\
                                  \vdots & \vdots & \vdots \\
                                  \Lambda_{m1}(t) & \cdots  & \Lambda_{mm}(t)
                                \end{array}
 \right],
\end{equation}
with operator entries $\Lambda_{jk}$ on the Fock space $\mathcal{F}$. Finally in this paper it is assumed that these quantum stochastic processes are \emph{canonical}, that is, they have the following non-zero Ito products
\begin{align}
 dB_j(t) dB_k^\ast(t) =& ~ \delta_{jk} dt, ~ d\Lambda_{jk}dB_l^\ast(t) = \delta_{kl}dB_j^\ast(t),  \label{Eq:CCR2}\\
 dB_j(t)d\Lambda_{kl}(t) =& ~ \delta_{jk}dB_l(t), ~ d\Lambda_{jk}(t)d\Lambda_{lm}(t) = \delta_{kl}d\Lambda_{jm}(t), ~ (j,k,l=1, \ldots, m) . \nonumber
\end{align}

\subsection{Open Quantum Systems in the $(S,L,H)$ Parametrization}\label{sec:open_systems}
\begin{figure}[tbh]
\centering
  \centering
  \includegraphics[width=2in]{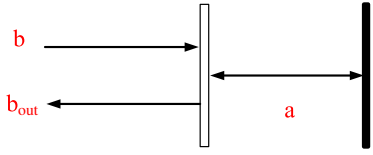}\\
  \caption{Open optical cavity, the white rectangle denotes a partially transmitting mirror at cavity resonant frequency}
  \label{fig_open_cavity}
\end{figure}
When a quantum system $G$ is driven by a Boson field $\mathcal{F}$, we have an \emph{open} quantum system. For example, if we allow the closed optical cavity in Fig.~\ref{fig_closed_cavity} to interact with a Boson field, we end up with an open optical cavity, Fig.~\ref{fig_open_cavity}.  While the mutual influence between the system and field may be described rigorously from first principles in terms of an interaction Hamiltonian, it is much more convenient to use an idealized quantum noise model which is valid under suitable rotating wave and Markovian assumptions, as in many situations in quantum optics, eg., cascaded open systems, see \cite{YD84,Gar93,Car93,YK03a} for detail. Let $\mathcal{A}_G$ and $A_{F}$ be physical variable spaces of the system $G$ and the field $\mathcal{F}$ respectively, then the physical variable space for the composite system is the tensor product space $\mathcal{A}_G \otimes A_{F}$.

Open quantum systems $G$ studied in this paper can be parameterized by a triple $(S,L,H)$ \cite{HP84,GJ09b}. Here, $S$ is a scattering matrix with entries in the system space $\mathcal{A}_G$, $L \in \mathcal{A}_G$ is an coupling operator that provides interface between systems and fields, $H \in \mathcal{A}_G$ is the internal Hamiltonian of quantum system $G$.

With these parameters, and assuming that the input field is canonical, that is, Eq. (\ref{Eq:CCR2}) holds, we have the following Schrodinger's equation for open quantum systems (in Ito form)
\begin{equation}\label{eq_U_linear}
dU(t) = \left\{ \mathrm{tr}[(S-I_m)d\Lambda^T] + dB^\dag(t)L - L^\dag S dB(t) - (iH+\frac{1}{2}L^\dag L)dt  \right\} U(t), ~ U(0) = I.
\end{equation}
(Note that for a closed system, Eq. (\ref{eq_U_linear}) becomes the familiar Schrodinger's equation (\ref{eq:U}).)
This, together with the evolution $X(t) = U(t)^\ast X U(t)$, yields the following quantum stochastic differential equations (QSDEs), in Ito form,
\begin{eqnarray}
 dX(t) &=& (-i[X(t),H(t)]+\mathcal{L}_{L(t)}(X(t)))dt+dB^\dag(t)S^\dag(t)[X(t),L(t)]+[L^\dag(t),X(t)]S(t)dB(t)  \label{eq_X} \\
       & & + \mathrm{tr}[(S^\dag(t)X(t)S(t)-X(t))d\Lambda^T(t)], ~~ X(0) = X, \nonumber
\end{eqnarray}
where the Lindblad  operator $\mathcal{L}_{L}$ is
\begin{equation} \label{landblad}
 \mathcal{L}_{L}(X) := \frac{1}{2}L^\dag[X,L]+\frac{1}{2}[L^\dag,X]L.
\end{equation}
For later use, we define a \emph{generator} operator
\begin{equation}\label{eq:generator}
\mathcal{G}_G(X) := -i[X,H] + \mathcal{L}_{L}(X).
\end{equation}
The output field $B_{out} (t) = U^\ast(t)B(t)U(t)$ satisfies
\begin{equation} \label{eq:output}
dB_{out}(t) = L(t)dt + S(t)dB(t).
\end{equation}
The gauge process of the output field $\Lambda_{out}(t) := \int_0^t b_{out}^\#(s)b_{out}^T(s)ds = U^\ast(t)\Lambda(t)U(t)$ satisfies
\begin{equation}
d\Lambda_{out}(t) = S^\#(t)d\Lambda(t)S^T(t) +  S^\#(t)dB^\#(t)L^T(t) + L^\#(t)dB^T(t)S^T(t) + L^\#(t)L^T(t)dt.
\end{equation}
Finally, in the Schrodinger picture, the reduced system density operator $\hat{\rho}$ satisfies the master equation, cf. \cite[Sec. 11.2.5]{GZ04},
\begin{equation}\label{eq:rho_2}
    \frac{d}{dt}\hat{\rho}(t) = -i[H, \hat{\rho}(t)] + \mathcal{L}^\prime_{L}(\hat{\rho}(t)) ,
\end{equation}
where the operator $\mathcal{L}^\prime_{L}$ is defined to be
\begin{equation}
\mathcal{L}^\prime_{L}(\hat{\rho}) := L^T\hat{\rho} L^\# - \frac{1}{2}L^\dag L\hat{\rho} -\frac{1}{2}\hat{\rho} L^\dag L.
\end{equation}

\begin{remark}{\rm
Clearly, open quantum systems presented in this section are quantum Markov processes.
}
\end{remark}

\subsection{Examples}

\subsubsection{Optical Cavity}\label{sec:cavity}
The one degree of freedom closed quantum harmonic oscillator in Fig. \ref{fig_closed_cavity} can be described by $(-,-,\omega a^\ast a)$, where the symbol ``-'' means that there is neither scattering nor coupling. The open optical cavity in Fig.~\ref{fig_open_cavity} may be described by $(1,\sqrt{\kappa}a, \omega a^\ast a)$, where $\kappa$ is coupling coefficient and $\omega$ is the resonant frequency. According to Eqs. (\ref{eq_X})-(\ref{eq:output}),
\begin{eqnarray}
  da(t) &=& -(i\omega + \frac{\kappa}{2})a(t)dt -\sqrt{\kappa}dB(t), ~ a(0) = a, \label{eq:cavity_1} \\
  dB_{out}(t) &=&  \sqrt{\kappa}a(t)dt + dB(t). \label{eq:cavity_2}
\end{eqnarray}

\subsubsection{Two-level Systems}\label{sec:two_level}
Given a two-level system parameterized by
\begin{equation*}
S_{-}=1, ~ L=\sqrt{\kappa }\sigma _{-}, ~ H=\frac{\omega}{2} \sigma _{z},
\end{equation*}%
with Pauli matrices
\begin{equation}\label{eq:pauli}
\sigma _{z}=\left[
\begin{array}{cc}
1 & 0 \\
0 & -1%
\end{array}%
\right], ~ \sigma _{-}=\left[
\begin{array}{cc}
0 & 0 \\
1 & 0%
\end{array}%
\right] , ~ \sigma _{+}=\left[
\begin{array}{cc}
0 & 1 \\
0 & 0%
\end{array}%
\right],~ \sigma _{x}=\left[
\begin{array}{cc}
0 & 1 \\
1 & 0%
\end{array}%
\right] , ~\sigma _{y}=\left[
\begin{array}{cc}
0 & -i \\
i & 0%
\end{array}%
\right],
\end{equation}
by Eq. (\ref{eq_U_linear}) the unitary operator $U(t)$ evolves according to
\begin{equation}
dU(t)=\left\{\sqrt{\kappa} dB^{\ast }(t)\sigma _{-}-\sqrt{\kappa}\sigma _{+}dB(t)-\frac{\kappa}{4}(\sigma_z+1)dt-i\frac{\omega}{2}\sigma_zdt\right\} U(t), ~ U(0) = I. \label{eq:U_2}
\end{equation}
 By Eqs. (\ref{eq_X}) and (\ref{landblad})
\begin{eqnarray}
  d\sigma_x(t) &=& -(\frac{\kappa}{2}\sigma_x(t)+ \omega\sigma_y(t))dt + \sqrt{\kappa}dB^\ast(t)\sigma_z(t)+ \sqrt{\kappa}\sigma_z(t)dB(t), \label{eq:sigma_x} \\
  d\sigma_y(t) &=& -(\frac{\kappa}{2}\sigma_y(t)-\omega\sigma_x(t))dt -i\sqrt{\kappa}dB^\ast(t)\sigma_z(t)+i\sqrt{\kappa}\sigma_z(t)dB(t), \label{eq:sigma_y}\\
  d\sigma_z(t) &=& -\kappa (I+\sigma_z(t))dt -\frac{\sqrt{\kappa}}{2}(\sigma_x(t)-i\sigma_y(t))dB^\ast(t)-\frac{\sqrt{\kappa}}{2}(\sigma_x(t)+i\sigma_y(t))dB(t). \label{eq:sigma_z}
\end{eqnarray}
On the other hand, the output field is%
\begin{equation}
dB_{out}(t)=\sqrt{\kappa}\sigma_-(t)dt+dB(t),  \label{eq:tla_bout}
\end{equation}%
\begin{equation}\label{eq:tla_gauge}
d\Lambda _{out}(t)=d\Lambda (t) + \sqrt{\kappa}dB^{\ast }(t)\sigma_-(t) + \sqrt{\kappa} \sigma_+(t)dB(t) + \frac{\kappa}{2}(\sigma_z+1) dt.
\end{equation}%

\begin{remark}
{\rm
It can be seen from Eq. (\ref{eq:sigma_x})-(\ref{eq:sigma_z}) that two-level systems are \emph{nonlinear} quantum systems.
}
\end{remark}

\section{Interconnection}\label{sec:interconnection}

In this section we discuss how two quantum systems can be connected to each other. More specifically we discuss concatenation product, series product, direct coupling, and linear fractional transform. Several examples from the literature are used to illustrate these interconnections. Propagation delays are ignored in interconnections. Discussions of influence of propagation delays on system performance can be found in, eg., \cite{IYY+11}.

\subsection{Concatenation Product}
\begin{figure}
\centering
  \includegraphics[width=1.5in]{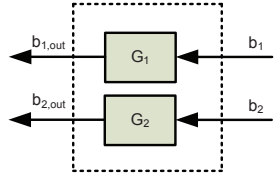}\\
  \caption{Concatenation product $G_1 \boxplus G_2$}
  \label{contatenation}
\end{figure}

 Given two open quantum systems $G_1 = (S_1,L_1,H_1)$ and  $G_2 = (S_2,L_2,H_2)$, their \emph{concatenation product}, Fig.~\ref{contatenation}, is defined to be
 \begin{equation}\label{eq:concatenation}
G_1 \boxplus G_2 := \left( \left[ \begin{array}{cc}
                                   S_1 & 0 \\
                                   0 & S_2
                                 \end{array}
 \right], \left[\begin{array}{c}
                  L_1 \\
                  L_2
                \end{array}
 \right],H_1+H_2 \right)
 \end{equation}

\subsection{Series Product}
\begin{figure}
\centering
  \includegraphics[width=3.5in]{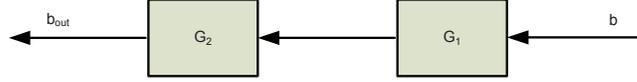}\\
  \caption{Series product $G_{2}\triangleleft G_{1}$}
  \label{cascade}
\end{figure}

 Given two open quantum systems  $G_1 = (S_1,L_1,H_1)$ and  $G_2 = (S_2,L_2,H_2)$ with the same number of input, their \emph{series product}, Fig.~\ref{cascade},  is defined to be
\begin{equation}\label{eq:series}
  G_2 \triangleleft G_1 := \left( S_2S_1, L_2+S_2L_1, H_1+H_2+\frac{1}{2i}(L_2^\dag S_2L_1-L_1^\dag S_2^\dag L_2) \right).
\end{equation}

\begin{theorem}
(Principle of Series Connections, \cite[Theorem 5.5]{GJ09b}): The parameters of the composite system $G_2  \leftarrow G_1$,
 obtained from $G_1 \boxplus G_2$
when the output of $G_1$ is used as input of $G_2$, is given by the series product $G_2 \triangleleft G_1$.
\end{theorem}

\subsection{Direct Coupling} \label{sec:models-direct}

\begin{figure}
\centering
  \includegraphics[width=0.7in]{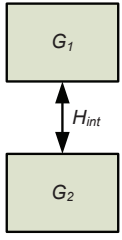}\\
  \caption{Directly coupled system $G_1 \bowtie G_2$}\label{direct_coupling}
\end{figure}

In quantum mechanics, two independent systems $G_1$ and $G_2$ may interact by exchanging energy. This energy exchange may be described by an {\em interaction Hamiltonian} $H_{int}$ of the form $H_{int} = X_1^\dagger X_2 + X_2^\dagger X_1$, where $X_1 \in \mathcal{A}_{G_1}$ and $X_2 \in \mathcal{A}_{G_2}$; see, eg., \cite{WM94b}, \cite{Lloyd00}, \cite{ZJ11}. In this case, we say the two systems $G_1$ and $G_2$ are \emph{directly coupled}, and the composite system is denoted $G_1 \bowtie G_2$, Fig.~\ref{direct_coupling}.

\subsection{Linear Fractional Transform}
\begin{figure}
\centering
  \includegraphics[width=2.5in]{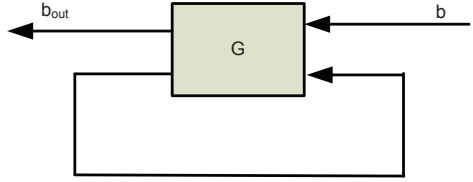}\\
  \caption{Linear fractional transformation $F(G)$}\label{network}
\end{figure}

Let $G$ in Fig.~\ref{network} be of the form $\left(\left[\begin{array}{cc}
                                                            S_{11} & S_{12} \\
                                                            S_{21} & S_{22}
                                                          \end{array}
\right], \left[\begin{array}{c}
                 L_1 \\
                 L_2
               \end{array}
 \right], H\right)$. Assume that $(I-S_{22})^{-1}$ exists. Then the conventional linear fractional transform yields a feedback network
 \[
 F(G) = (S_{11}+S_{12}(I-S_{22})^{-1}S_{21}, L_1+S_{12}(I-S_{22})^{-1}L_2, H+\mathrm{Im}\{L_1^\dag S_{12}(I-S_{22})^{-1}L_2\}+\mathrm{Im}\{L_2^\dag S_{22}(I-S_{22})^{-1}L_2\})
 \]
 from input $b$ to output $b_{out}$.

\subsection{Examples} \label{sec:examples}
In this section examples in the literature are used to demonstrate the usefulness of the parametrization $(S,L,H)$ and interconnections. More examples can be found in \cite{WM94b,GJ09b,Gough08,Mabuchi11}. For the convenience of the readers to refer to the original papers we use symbols in those original papers.

\subsubsection{Example 1 (Carmichael (1993))} \label{sec:Car93}

In \cite{Car93} quantum trajectory theory is formulated for interaction of open quantum systems via series product, Fig.~\ref{cascade}.
Given two open systems $G_1 = (1,L_A,H_A),G_2 = (1,L_B,H_B)$ with $L_A = \sqrt{2\kappa_A}a_A,  L_B = \sqrt{2\kappa_B}a_B$. Here $\kappa_A$ and $\kappa_B$ are coupling constants and $a_A$ and $a_B$ are annihilation operators for systems $G_1$ and $G_2$ respectively. The series product yields
\begin{equation}
    G_2 \vartriangleleft G_1 =(1,\sqrt{2\kappa_A}a_A+\sqrt{2\kappa_B}a_B,H_A+H_B+i\sqrt{\kappa_A\kappa_B}(a_A^\ast a_B- a_B^\ast a_A)).
\end{equation}
Identifying $H_A+H_B+i\sqrt{\kappa_A\kappa_B}(a_A^\ast a_B- a_B^\ast a_A))$ with $\hat{H}_S$ in \cite[Eq.(7)]{Car93} and $\sqrt{2\kappa_A}a_A+\sqrt{2\kappa_B}a_B$ with $\hat{C}$ in \cite[Eq.(9)]{Car93} respectively, Eq. (\ref{eq:rho_2}) re-produces the master equation \cite[Eq.(8)]{Car93}.

\subsubsection{Example 2 (Gardiner (1993))} \label{sec:Gar93}

\begin{figure}
\centering
  \includegraphics[width=3.5in]{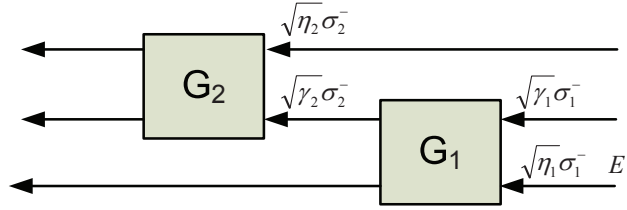}\\
  \caption{A re-draw of \cite[FIG.1]{Gar93}. For clarity, coupling operators $\sqrt{\gamma_1}\sigma_1^-, \sqrt{\eta_1}\sigma_1^-, \sqrt{\gamma_1}\sigma_2^-,\sqrt{\eta_2}\sigma_1^-$ are shown explicitly. $E$ is a coherent electric drive of $G_1$.}\label{fig_Gar93}
\end{figure}

In \cite{Gar93} quantum Langevin equations and a quantum master equation were derived for a cascade of two two-level systems, Fig.~\ref{fig_Gar93}. The two two-level systems $G_1$ and $G_2$ in  Fig.~\ref{fig_Gar93} are respectively
\[
G_1 = (1,\sqrt{\gamma_1}\sigma_1^-,0) \boxplus ((1,\sqrt{\eta_1}\sigma_1^-,0)\vartriangleleft (1,E,0)),
\]
\[
G_2 = (1,\sqrt{\eta_2}\sigma_2^-,0) \boxplus (1,\sqrt{\gamma_2}\sigma_2^-,0).
\]
Here, $\gamma_1$ and $\eta_1$ are coupling constants for system $G_1$, $E$ is an incident coherent electric drive (not an operator) of $G_1$.  $\gamma_1$ and $\eta_1$ are coupling constants for system $G_2$. $\sigma^-$ is defined in Eq. (\ref{eq:pauli}):
\[
\sigma_1^- = \sigma_2^- = \left[ \begin{array}{cc}
                                   0 & 0 \\
                                   1 & 0
                                 \end{array}
  \right].
\]
Let $G$ be a series product $ G = (G_2 \boxplus (1,0,0)) \vartriangleleft ((1,0,0)\boxplus G_1)$. With these, according to Eq. (\ref{eq:rho_2}), \cite[Eq.(14)]{Gar93} can be re-produced. (Notice the fact that in the interaction picture $H_{sys}=0$ is used.)

\subsubsection{Example 3 (Sherson and Molmer (2006); Sarma, Silberfarb, and  Mabuchi (2008))}

A scheme is proposed in \cite{SM06} to produce continuous-wave fields or pulses of polarization-squeezed light by passing classical, linearly polarized laser light through an atomic sample twice; that is, the field output of the first pass is fed back to the atomic sample again so as to generate polarization-squeezed light. This scheme is confirmed and extended in \cite{SSM08}. The atomic sample can be modeled as an open quantum system $G = G_1 \boxplus G_2$ with
\[
G_1 = (1, \frac{1}{\sqrt{2}}\alpha p,0), ~~ G_2 = (1, -\frac{i}{\sqrt{2}}\alpha x,0).
\]
Here $\alpha$ is a real constant, and $x,p$ are position and momentum operators respectively, cf. \cite[Eqs. (A1)-(A2)]{SSM08}. Double-pass of an electromagnetic field through the atomic sample introduces a series product, that is the overall quantum system is
\[
G_2 \vartriangleleft G_1 = (1,\frac{\alpha}{2}(p-ix),\frac{\alpha^2}{4}(xp+px)),
\]
whose Schrodinger equation is \cite[Eq.(1)]{SSM08}.


\subsubsection{Example 4 (Zhang and James (2011))} \label{sec:direct_coupling_example}

Given two closed quantum Harmonic oscillators $G_1$ and $G_2$ as studied in Section \ref{sec:closed_qho},  we take the interaction Hamiltonian $H_{int}$ in Fig.~\ref{direct_coupling} to be
\begin{eqnarray}
H_{int}
=  \frac{1}{2} \left(  \breve{a}^{(1) \dagger} \Xi^\dagger  \breve{a}^{(2)} + \breve{a}^{(2) \dagger} \Xi   \breve{a}^{(1)}
\right),
\label{eq:direct-1}
\end{eqnarray}
where $ \Xi = \Delta( i K_-,i K_+) $ for matrices $K_{-}, K_{+} \in \mathbb{C}^{n_{2}\times n_{1}}$. The Hamiltonian for the directly coupled  system $G_1 \bowtie G_2$ is
\begin{equation}
H = H_{0,1} + H_{int} +H_{0,2} ,
\label{eq:direct-3}
\end{equation}
where $H_{0,k} = \frac{1}{2} \breve{a}^{(k)\dagger} \Delta( \Omega^{(k)}_-, \Omega^{(k)}_+) \breve{a}^{(k)}$ is the self-Hamiltonian for $G_k$, and $H_{int}$ is given by (\ref{eq:direct-1}). It is easy to show that  system operators $\breve{a}^{(j)}(t) = U^\ast(t) \breve{a}^{(j)} U(t)$ ($j=1,2$) satisfy the following linear differential equations, in Stratonovich form,
\begin{eqnarray*}
\dot{\breve{a}}^{(1)}(t)  &=& A_{0,1} \breve{a}^{(1)} (t) + B_{12} \breve{a}^{(2)}(t) ,  ~~ \breve{a}^{(1)}(0)=\breve{a}^{(1)} , \\
\dot{\breve{a}}^{(2)}(t)  &=& A_{0,2} \breve{a}^{(2)} (t) + B_{21} \breve{a}^{(1)}(t) ,  ~~ \breve{a}^{(2)}(0)=\breve{a}^{(2)} ,
\end{eqnarray*}
where
\[
A_{0,j}=-\Delta(i \Omega^{(j)}_-, i \Omega^{(j)}_+), ~ B_{12} = - \Delta( K_-, K_+)^\flat, ~ B_{21} = - B_{12}^\flat , ~  (j=1,2).
\]

\section{Quantum Dissipative Systems}\label{sec:general_sytemes}

\begin{figure}
\centering
  \includegraphics[width=1.5in]{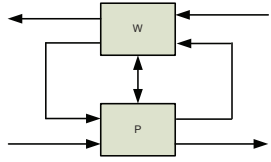}\\
  \caption{Plant-exosystem network $P\wedge W$}\label{feedback}
\end{figure}

Open systems are systems that interact with other systems and/or their environment. In classical control theory, a general framework for the stability of open systems has been developed \cite{Wil72,vdS96,WT02,Wil07}. This classical theory abstracts energy concepts and provides fundamental relations for stability in terms of generalized energy inequalities. In this section we briefly review dissipation theory for open quantum systems,   \cite{JG10}. In Fig.~\ref{feedback} the open quantum system $P=(S,L,H)$ is the plant of interest whose space of variables is denoted $\mathcal{A}_P$. The other open quantum system $W=(R,w,D)$ is an external system or the environment, whose space of variables is denoted $\mathcal{A}_{ex}$. $W$ is called an \emph{exosystem}. Moreover, we allow $W$ to vary in a class of such exosystems $\mathcal{W}$.

\subsection{Dissipativity, Stability, Passivity, Gain}

In this section we present concepts of dissipativity, stability, passivity and gain. As with the classical case, criteria of stability, passivity and gain follow those of dissipativity.

The following assumption is used in the sequel.

\noindent \emph{Assumption A1.}  The inputs to the composite system $P\wedge W$ are all canonical vacuum fields, cf. Section \ref{sec:field}.

Let $r_P(W)$ be a self-adjoint operator in the composite plant-exosystem space $\mathcal{A}_P \otimes \mathcal{A}_{ex}$. $r_P(W)$ is usually called \emph{supply rate}. We have the following definition of dissipativity for open quantum systems $P$.

\begin{definition}(Dissipation, \cite[Sec. III-A]{JG10}) \label{def:dissipativity}
The plant $P$ is said to be \emph{dissipative} with supply rate $r_P(W)$ with respect to a class of exosystems $\mathcal{W}$ if there exists a non-negative plant observable $V\in \mathcal{A}_P$, called \emph{storage function}, such that the dissipation inequality
\begin{equation}\label{eq:dissipation_def}
\underbrace{\mathbb{E}_0\left[ V(t)-V(0)\right]}_{\mathrm{stored~ energy}} \leq \underbrace{\int_0^t \mathbb{E}_0\left[ r_P(W)(s) \right]ds}_{\mathrm{supplied ~ energy}}
\end{equation}
holds for all $W\in \mathcal{W}$ and all $t\geq0$, where $\mathbb{E}_0$ is vacuum expectation \cite[Chapter 26]{Pa92}. In particular, when ``='' in (\ref{eq:dissipation_def}) holds for all $W\in \mathcal{W}$ and all $t\geq0$, $P$ is called \emph{lossless}.
\end{definition}

The combination of Definition \ref{def:dissipativity} and the following property of vacuum expectation \cite[Proposition 26.6]{Pa92}
\begin{equation}\label{eq:vacuum_exp}
\mathbb{E}_\tau[V(t)] = V(\tau) +\int_\tau^t \mathbb{E}_0[\mathcal{G}_{P\wedge W}(V(r))]dr,
\end{equation}
yields an infinitesimal version (namely independent of the time variable) of the dissipation inequality (\ref{eq:dissipation_def}).

\begin{theorem} \label{thm:passipation} (Dissipation, \cite[Theorem 3.1]{JG10})
Pertaining to Fig.~\ref{feedback}, the plant $P$ is dissipative with a supply rate $r_P(W)$ with respect to a class of exosystems $\mathcal{W}$ if and only if there exists a non-negative plant observable $V\in \mathcal{A}_P$ such that
\begin{equation}\label{eq:dissipation}
\mathcal{G}_{P\wedge W}(V) - r_P(W) \leq 0
\end{equation}
holds for all $W\in \mathcal{W}$.
\end{theorem}

Given a quantum system $P$, assume it has indirect and/or direct connections to external systems and/or its environment. With slight abuse of notation we still call $P$ an open quantum system. Let $\mathcal{W}_u$ denote the class of all the quantum systems that can be connected to $P$, directly or indirectly. The following result shows that $P$ is lossless with respect to $\mathcal{W}_u$.

\begin{theorem}(\cite[Theorem 3.3]{JG10}) \label{thm:universal_dissipativity}
Pertaining to Fig.~\ref{feedback}, for any given storage function $V_0$, which is a non-negative observable in $\mathcal{A}_P$, the open quantum system $P$ is lossless with a supply rate
\begin{equation}\label{eq:natural_supply_rate}
r_0(W) = \mathcal{G}_{P\wedge W}(V_0)
\end{equation}
 with respect to $\mathcal{W}_u$.
\end{theorem}

Theorem \ref{thm:universal_dissipativity} shows that any open quantum system is dissipative in some sense.

Next we study stability of open quantum systems which is characterized in terms of the evolution of mean values.

\begin{definition}(Exponential stability, \cite[Sec. III-B]{JG10}) \label{def:stability}
An open quantum system $P$ is said to be \emph{exponentially stable} if there exists a non-negative observable $V\in \mathcal{A}_P$,  scalars $c>0$ and $\lambda \geq 0$ such that
\begin{equation}
\langle V(t) \rangle \leq e^{-ct}\langle V\rangle +\frac{\lambda}{c}
\end{equation}
holds for any plant state and all time $t \geq 0$. Moreover, if $\lambda = 0$, then $\lim_{t\to\infty}\langle V(t) \rangle = 0$.
\end{definition}

The combination of Eq. (\ref{eq:vacuum_exp}) and Definition \ref{def:stability} gives the following stability result for open quantum systems $P$.

\begin{theorem}\label{thm:stability}(Stability, \cite[Lemma 3.4]{JG10})
If there exists a nonnegative observable $V \in \mathcal{A}_P$, scalars $c>0$ and $\lambda \geq 0$ such that
 \begin{equation*}
    \mathcal{G}_P(V) + c V \leq \lambda,
 \end{equation*}
 then the open quantum system $P$ is exponentially stable. Moreover, if $\lambda = 0$, then $\lim_{t\to\infty}\langle V(t) \rangle = 0$.
\end{theorem}

\begin{figure}
\centering
  \centering
  \includegraphics[width=3in]{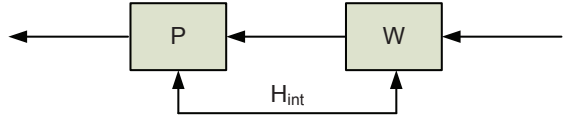}\\
  \caption{Series product plus direct coupling}\label{cd}
\end{figure}

In what follows we focus on the series product of $P$ and $W$ with additional direct coupling, Fig.~\ref{cd}. That is, the composite system is
\begin{equation*}
    P\wedge W = (P\triangleleft W) \boxplus (-,-,H_{int}) ,
\end{equation*}
where $P=(I,L,H)$, $W = (I, w, 0)$, $H_{int} = -i(M^\dag v-v^\dag M)$ with $w,v\in  \mathcal{A}_{ex}$ and $M\in \mathcal{A}_P$. With slight abuse of notation, we write
\begin{equation*}
    P\wedge W = P\triangleleft W,
\end{equation*}
where $W = (I, w, -i(M^\dag v-v^\dag M))$. That is, direct coupling is absorbed into the exosystem $W$.

Let $V \in \mathcal{A}_P$ be a non-negative observable. Assume that $S=I$. By Eq. (\ref{eq:generator})
\begin{equation}\label{eq:natural_supply_rate_2}
\mathcal{G}_{P\wedge W}(V) = \mathcal{G}_{P}(V) + \mathcal{L}_w(V) + [w^\dag ~~ v^\dag]Z + Z^\dag \left[\begin{array}{c}
                                                                                                                   w \\
                                                                                                                   v
                                                                                                                 \end{array}
 \right] + [V,v^\dag]M - M^\dag [V,v]
\end{equation}
where $Z = \left[V,\left[\begin{array}{c}
                        L \\
                        M
                      \end{array}
\right]\right]$.

For fixed $M\in \mathcal{A}_P$, define a class of exosystems
\begin{equation}\label{eq:W1}
\mathcal{W}_1 = \{W=(I,w,-i(M^\dag v-v^\dag M): w,v \mathrm{~commute ~ with~} \mathcal{A}_P \}.
\end{equation}
Then we have the following definition of passivity for the system $P$ in Fig.~\ref{cd}.

\begin{definition}\label{def:passivity} (Passivity, \cite[Sec. III-C]{JG10})
Given $M\in \mathcal{A}_P$, the plant $P=(I,L,H)$ is said to be \emph{passive} with respect to the class of exosystems $\mathcal{W}_1$ in (\ref{eq:W1}) if it is dissipative with the supply rate
\begin{equation}
r_P(W) = -N^\dag N +[w^\dag ~~ v^\dag]Z + Z^\dag \left[ \begin{array}{c}
                                                          w \\
                                                          v
                                                        \end{array}
 \right]+\lambda
\end{equation}
for some non-negative real number $\lambda$, and $N,Z\in \mathcal{A}_P$. $P$ is said to be strictly passive if $N^\dag N$ is strictly positive.
\end{definition}

The combination of Definition \ref{def:passivity}, Theorem \ref{thm:passipation} and Eq. (\ref{eq:natural_supply_rate_2}) gives the following passivity result.

\begin{theorem}(Positive Real Lemma, \cite[Theorem 3.6]{JG10})\label{thm:prl}
A plant  $P=(I,L,H)$ is passive with respect to the class of exosystems $\mathcal{W}_1$ in (\ref{eq:W1}) if and only if there exists a non-negative observable $V\in \mathcal{A}_P$, an operator $N\in \mathcal{A}_P$, and a non-negative real number $\lambda$ such that
\begin{eqnarray}
 \mathcal{ G}_P(V) + N^\dag N - \lambda &\leq & 0, \label{eq:prl-2} \\
  Z &=& \left[V,\left[\begin{array}{c}
                L \\
                M
              \end{array}\right]
  \right]. \label{eq:prl-5}
\end{eqnarray}
\end{theorem}

As with the classical case, strict passivity implies stability.

\begin{theorem} \label{thm:strict_passivity_stability}
The open quantum system $P$ is exponentially stable if it is strictly passive with respect to the exosystem $W=(I,0,0)$.
\end{theorem}

The bounded real lemma is used to determine $L^2$ gain of an open system, and in conjunction with the small gain theorem, can be used for robust stability analysis and design. In what follows we discuss $L^2$ gain of open quantum systems.

Define a class of exosystems
\begin{equation}\label{eq:W2}
\mathcal{W}_2 = \{W=(I,w,0): w \mathrm{~commutes ~ with~} \mathcal{A}_P\}.
\end{equation}
(Note that in this case there is no direct coupling.)

\begin{definition}\label{def:gain}($L^2$ gain, \cite[Sec. III-C]{JG10})
The plant $P=(I,L,H)$ is said to have $L^2$ gain $g>0$  with respect to the class of exosystems $\mathcal{W}_2$ in (\ref{eq:W2}) if it is dissipative with the supply rate
\begin{equation}
r_P(W) = g^2 w^\dag w - (N+Zw)^\dag (N+Zw) +\lambda
\end{equation}
for some non-negative real number $\lambda$, and $N,Z\in \mathcal{A}_P$.
\end{definition}

The combination of Definition \ref{def:gain} and Theorem \ref{thm:passipation} gives the following result.

\begin{theorem}\label{thm:BRL_0} (Bounded Real Lemma, \cite[Theorem 3.7]{JG10})
A plant $P=(I,L,H)$ has $L^2$ gain $g>0$ with respect to $\mathcal{W}_2$ if and only if there exists a non-negative plant variable $V\in \mathcal{A}_P$, an operator $N\in \mathcal{A}_P$, and a non-negative real number $\lambda$ such that
\begin{equation}
    \Gamma = g^2 -Z^\dag Z \geq 0
\end{equation}
and
\begin{equation}
    \mathcal{G}_P(V) + N^\dag N -w^\dag \Gamma w + w^\dag ([V,L]+Z^\dag N)+([V,L]+Z^\dag N)^\dag w -\lambda \leq 0
\end{equation}
for all $w\in \mathcal{A}_{ex}$. If $\Gamma^{-1}$ exists, then plant $P=(I,L,H)$ has gain $g>0$ with respect to $\mathcal{W}_2$ if and only if
\begin{equation}
 \mathcal{G}_P(V) + N^\dag N + ([V,L]+Z^\dag N)^\dag\Gamma^{-1}([V,L]+Z^\dag N) -\lambda \leq 0
\end{equation}
for all $w\in \mathcal{A}_{ex}$. In the latter case, the plant $P$ is strictly bounded real.
\end{theorem}

\subsection{Example} \label{sec:general_example}
The following example illustrates the above results for stability, passivity, and $L^2$ gain. Consider a two-level system $P$ and an exosystem $W$ of the form
\[
P=(1,\sqrt{\gamma}\sigma_+, \frac{\omega}{2}\sigma_z), ~~  W = (1, w,0),
\]
where $w$ commutes with $\mathcal{A}_P$. Assume there is no direct coupling between $P$ and $W$. Choose a storage function $V_0 = \frac{1}{2}(I-\sigma_z) = \sigma_-\sigma_+$ and a supply rate
\begin{equation}\label{eq:supply_rate}
r_P(W) =\mathcal{G}_{P\wedge W}(V_0) = \mathcal{G}_{P}(V_0) + w^\ast Z + Z^\ast w,
\end{equation}
where $Z = [V_0,\sqrt{\gamma}\sigma_+]=-\sqrt{\gamma}\sigma_+$. Clearly, $\mathcal{G}_{P}(V_0) = -\gamma V_0$. As a result, Eq. (\ref{eq:supply_rate}) becomes
\begin{eqnarray*}
  r_P(W) &=& -\gamma V_0 -\sqrt{\gamma}(w^\ast \sigma_+ + \sigma_- w) \\
         &=& -(\sqrt{\gamma}\sigma_++w)^\ast (\sqrt{\gamma}\sigma_++w) + w^\ast w .
\end{eqnarray*}

Choose $N = \sqrt{\gamma}\sigma_+$ and $Z=-N$, by Theorem \ref{thm:prl} we see that the system is passive. Choose $N = \sqrt{\gamma}\sigma_+$ and $Z=1$, by Theorem \ref{thm:BRL_0} we find that the system has $L^2$ gain 1. Finally, when $W=(1,0,0)$, $r_P(W) = r_P(I)= -\gamma V_0$, then by Theorem \ref{thm:stability} the system $P$ is exponentially stable.

\section{Linear Quantum Systems}\label{sec:models-general}

Linear quantum systems are those for which certain conjugate operators evolve linearly, the optical cavity being a basic example, cf. Section \ref{sec:cavity}. Linear systems have the advantage that they are much more computationally tractable than general nonlinear systems, and indeed, powerful methods from linear algebra may be exploited.

\subsection{General Model}\label{sec:models}

\begin{figure}
\centering
  \includegraphics[width=3in]{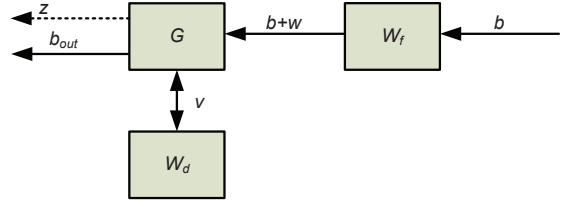}\\
  \caption{General model}\label{general_model}
\end{figure}

Open linear quantum systems discussed in this paper are open quantum harmonic oscillators with direct and indirect couplings to other quantum systems and/or external fields. In this section we present a general model for an open linear quantum system $G$, Fig.~\ref{general_model}, based on the ingredients discussed in the previous sections. Here, $G$ is an open quantum system with parametrization $(I,L,H_0)$, where $L= C_-a+C_+a^\#$ with $C_-,C_+$ being constant complex-valued matrices. The internal Hamiltonian $H_0$ is that given in Eq. (\ref{eq:H0}). Moreover, $G$ is allowed to coupled directly to another (independent) quantum system $W_d$ via an interaction Hamiltonian
\begin{equation}
H_{int}
=  \frac{1}{2} \left(     \breve{a}^{ \dagger} \Xi^\dagger  \breve{v} + \breve{v}^\dagger \Xi   \breve{a}
\right) ,
\label{eq:gen-3}
\end{equation}
where $\Xi=\Delta(i K_-, i K_+)$. Our interest is in the influence of external systems/fields on the given system $G$.  The performance characteristics of interest are encoded in a {\em performance variable}\footnote{A performance variable is chosen to capture control performance, such as an error quantity, and  so may involve external variables, like a reference signal. Performance variables need not have anything to do with the output quantities associated with direct or indirect couplings to other systems, cf. Sections \ref{sec:performance_specification} and \ref{sec:feedback}.}\, $z$.

Building upon the discussions in previous sections, the equations for $G$ (including direct coupling, indirect coupling and performance variable) are
\begin{eqnarray}
\dot{\breve{a}}(t) & \hspace{-2mm}  =& \hspace{-2mm} A \breve{a}(t) + B_d \breve{v}(t) + B_f \breve{w}(t) + B_f \breve{b}(t), ~ \breve{a}(0) = \breve{a}, \label{eq:gen-1-dyn} \\
\breve b_{out}(t) & \hspace{-2mm}  =& \hspace{-2mm} C_f \breve{a}(t) +  \breve{w}(t) + \breve{b}(t),
\label{eq:gen-1-f}  \\
\breve{z}(t) & \hspace{-2mm}  =& \hspace{-2mm} C_p \breve{a}(t) + D_{pd} \breve{v}(t) + D_{pf}  \breve{w}(t),
\label{eq:gen-1-p}
\end{eqnarray}
The complex matrices in (\ref{eq:gen-1-dyn}) and  (\ref{eq:gen-1-f}) are given by
\begin{equation}
 A=-\frac{1}{2}C_f^{\flat}C_f-\Delta\left(i\Omega_{-},i\Omega_{+}\right), ~ B_d = -\Delta( K_-, K_+)^\flat,
\label{eq:gen-2}
\end{equation}
\begin{equation}
C_f = \Delta(C_-, C_+), ~~ B_f = - C_f^\flat .
\label{eq:gen-2b}
\end{equation}
The matrices $A$ and $B_f$  are specified by the parameters $\Omega_\pm$, and $C_\pm$. In equations  (\ref{eq:gen-1-dyn}) and  (\ref{eq:gen-1-f}), $b(t)$ and $b_{out}(t)$ are respectively, the input and output fields for $G$. The term $v$ in (\ref{eq:gen-1-dyn}) is an exogenous quantity associated with  $W_d$ with which $G$ is directly coupled via the interaction Hamiltonian $H_{int}$, cf. Section \ref{sec:direct_coupling_example}.

The term $w$ in (\ref{eq:gen-1-dyn})  is another  exogenous quantity associated with another (independent) system $W_f$ with which $G$ is indirectly coupled through a series product. $W_f$ may be a quantum system of the form $(1,w,0)$ where $w$ is an operator on some Fock space, it can also denote modulation so that $w$ coherent drive modulates the vacuum field $b$ cf. $E$ in Example 2 of Section \ref{sec:examples}. Because of the assumed independence, $w$ and $v$ commute with the mode operators $a_j, a_j^\ast$ for $G$. While $v$ and $w$ are arbitrary external variables, the time evolutions $v(t)$ and $w(t)$ (when it is an operator) are determined by the evolution of the overall composite system. The matrices $C_p$, $D_{pd}$ and $D_{pf}$ specify the performance variable $z$. In brief, system $G$ is specified by the parameters $G = (\Omega_\pm, C_\pm, K_\pm, C_p, D_{pd}, D_{pf})$. Of these, $\Omega_\pm$,  $C_\pm$ and $K_\pm$ are {\em physical} parameters.

In particular, when all the plus terms are zero, namely $C_+ = 0, \Omega_+ = 0, K_+=0$, all matrices $A,B_f,B_d,C_f$ are block diagonal, system (\ref{eq:gen-1-dyn})-(\ref{eq:gen-1-f}) is equivalent to
\begin{eqnarray}
  \dot{a}(t) &=& -(i\Omega_- + \frac{1}{2}C_-^\dag C_- )a(t) - K_-^\dag v(t) - C_-^\dag w(t) - C_-^\dag b(t), ~ a(0)=a, \label{eq:system_passive_a} \\
  b_{out}(t) &=& C_- a(t) + b(t),  \label{eq:system_passive_b}
\end{eqnarray}
cf. optical cavity (\ref{eq:cavity_1})-(\ref{eq:cavity_2}). It can be readily shown that system (\ref{eq:system_passive_a})-(\ref{eq:system_passive_b}) is \emph{passive}. Passive systems have been studied in, eg., \cite{MP11,MP11a,petersen11,ZJ11,ZJ11b}.

\subsection{Physical Realizability}
It can be readily verified that the following relations for system matrices (\ref{eq:gen-2})-(\ref{eq:gen-2b}) hold
\begin{eqnarray}
J_n A + A^\dagger  J_n  + C_f^\dagger    J_{m} C_f &=& 0 , \label{eq:c-2-ccr} \\
B_f &=& -C^\flat_f , \label{eq:c-2-f} \\
B_d &=& -\Delta( K_-, K_+)^\flat . \label{eq:c-2-d}
\end{eqnarray}
Equation (\ref{eq:c-2-ccr}) characterizes preservation of the canonical commutation relations, namely
\begin{equation}
[ \breve{a}_j(t), \breve{a}_k^\ast(t) ] = [ \breve{a}_j, \breve{a}_k^\ast ] = (J_n)_{jk}, ~ \forall t\geq 0,  ~~ (j,k=1,\ldots, n).
\label{eq:c-0}
\end{equation}
Eq. (\ref{eq:c-2-f}) reflects the input and output relation, while Eq. (\ref{eq:c-2-d}) is for direct coupling.

The relations (\ref{eq:c-2-ccr})-(\ref{eq:c-2-d}) are called \emph{physical realizability} relations, which generalize results in \cite[Theorem 3.4]{JNP08}, \cite{SP09}, \cite[Theorem 5.1]{MP11a}, \cite[Theorem 3]{SP09}. These conditions guarantee that the equations correspond to a physical system.

\subsection{Quadrature Representation}\label{sec:models-quadrature}

So far, {\em annihilation-creation representation} has been used to represent linear quantum systems in terms of the notation $\breve{a} = [a^T ~ a^\dagger]^T$, the resulting matrices are complex-valued matrices. In this section we introduce an alternative representation, the so-called {\em quadrature representation}, which leads to equations with real-valued matrices.

Define the unitary matrix
\begin{equation}
\Lambda  = \frac{1}{\sqrt{2}} \left[
\begin{array}{cc}
I & I
\\
-i I & i I
\end{array}
\right]
\label{eq:quad-1}
\end{equation}
and the vector of self-adjoint operators
\begin{equation}
\tilde  a  =   \left[
\begin{array}{c}
q
\\
p
\end{array}
\right]
\label{eq:quad-2}
\end{equation}
by the relation
\begin{equation}
\tilde a = \Lambda  \breve{a}.
\label{eq:quad-3}
\end{equation}
 The vector $q = \frac{1}{\sqrt{2}} [I ~ I] \breve{a}$ is known as the {\em real quadrature}, while $p = \frac{1}{\sqrt{2}} [-i I ~ i I] \breve{a}$ is called the {\em imaginary} or {\em phase quadrature} \cite{WM08}.

Similarly define unitary matrices $\Lambda_f$, $\Lambda_d$ and $\Lambda_p$ of suitable dimension,  of the form (\ref{eq:quad-1}), and define quadrature vectors
\[
\tilde b = \Lambda_f  \breve{b}, ~ \tilde b_{out} = \Lambda_f  \breve{b}_{out}, ~
\tilde w = \Lambda_f  \breve{w}, ~ \tilde v = \Lambda_d  \breve{v}, ~ \tilde z = \Lambda_p \breve z .
\]
Then in quadrature form $G$ is in the form
\begin{eqnarray}
\dot{\tilde{a}}(t) &=&  \tilde  A \tilde{a}(t) +  \tilde  B_d \tilde{v}(t) +  \tilde  B_f \tilde{w}(t) + \tilde  B_f \tilde{b}(t), ~ \tilde{a}(0) = \tilde{a},  \label{eq:qradrature} \\
\tilde{b}_{out}(t) &=&
 \tilde  C_f \tilde{a}(t) +  \tilde{w}(t) + \tilde{b}(t), \nonumber  \\
\tilde{z}(t) &=&  \tilde  C_p \tilde{a}(t) + \tilde  D_{pd} \tilde{v}(t) + \tilde D_{pf}  \tilde{w}(t), \nonumber
\end{eqnarray}
where $\tilde A = \Lambda A \Lambda^\dagger$,
 $\tilde B_d  = \Lambda B_d  \Lambda^\dagger_d$,
$\tilde B_f  = \Lambda B_f  \Lambda^\dagger_f$,
 $\tilde C_f  = \Lambda_f C_f  \Lambda^\dagger$,
$\tilde C_p  = \Lambda_p C_p  \Lambda^\dagger$,
$\tilde D_{pd}  = \Lambda_p D_{pd}   \Lambda^\dagger_d$,
$\tilde D_{pf}  = \Lambda_p D_{pf}   \Lambda^\dagger_f$.
 Note that all entries of the matrices in this representation are real.

\subsection{Series Products for Linear Quantum Systems}\label{sec:series_linear}

Assume both $G_1$ and $G_2$ in Fig.~\ref{cascade} are linear, in this section we present the explicit from of $G = G_2 \triangleleft G_1$.

For ease of presentation we assume both $G_1$ and $G_2$ are passive with parametrization $G_j = (I, C_-^{(j)}a^{(j)}, 0)$, ($j=1,2$). Therefore
\begin{eqnarray}
  \dot{a}^{(j)}(t) &=& -\frac{1}{2} (C_-^{(j)})^\dag C_-^{(j)} a^{(j)}(t) - (C_-^{(j)})^\dag b^{(j)}(t),  ~ a^{(j)}(0) = a^{(j)}, \label{eq:system_passive_c}\\
  b_{out}^{(j)}(t) &=&  C_-^{(j)} a^{(j)}(t) + b^{(j)}(t), ~~ (j=1,2).  \label{eq:system_passive_d}
\end{eqnarray}
According to Eq. (\ref{eq:series}), the composite linear quantum system $G$ is
 \begin{equation}\label{linear_system_series}
 G = \left(I, [C_-^{(1)} ~  C_-^{(2)}]\left[\begin{array}{c}
                                                  a^{(1)} \\
                                                  a^{(2)}
                                                \end{array}
  \right], \frac{1}{2i}[(a^{(1)})^\dag  ~   (a^{(2)})^\dag]\left[\begin{array}{cc}
                                                                 0 & -(C_-^{(1)})^\dag C_-^{(2)} \\
                                                                 (C_-^{(2)})^\dag C_-^{(1)} & 0
                                                               \end{array}
   \right]\left[\begin{array}{c}
                                                  a^{(1)} \\
                                                  a^{(2)}
                                                \end{array}
  \right]\right).
 \end{equation}
 By Eqs. (\ref{eq:system_passive_a})-(\ref{eq:system_passive_b}), $G$ is in the form of
 \begin{eqnarray}
   \left[ \begin{array}{c}
            \dot{a}^{(1)}(t) \\
            \dot{a}^{(2)}(t)
          \end{array}
    \right] &=& - \left[ \begin{array}{cc}
                           \frac{1}{2}(C_-^{(1)})^\dag C_-^{(1)} & 0 \\
                           (C_-^{(2)})^\dag C_-^{(1)} & (C_-^{(2)})^\dag C_-^{(2)}
                         \end{array}
     \right]\left[ \begin{array}{c}
            a^{(1)}(t) \\
            a^{(2)}(t)
          \end{array}
    \right] - \left[\begin{array}{c}
                      (C_-^{(1)})^\dag \\
                      -(C_-^{(2)})^\dag
                    \end{array}
     \right]b^{(1)}(t),  \label{linear_system_series_a} \\
   b_{out}^{(2)}(t) &=&  [C_-^{(1)} ~  C_-^{(2)}]\left[\begin{array}{c}
                                                  a^{(1)}(t) \\
                                                  a^{(2)}(t)
                                                \end{array}
  \right] + b^{(1)}(t) .  \label{linear_system_series_b}
 \end{eqnarray}

If we identify $b^{(2)}(t)$ with $b_{out}^{(1)}(t)$ in Eqs. (\ref{eq:system_passive_c})-(\ref{eq:system_passive_d}),  then Eqs. (\ref{eq:system_passive_c})-(\ref{eq:system_passive_d}) give rise to system (\ref{linear_system_series_a})-(\ref{linear_system_series_b}) too. This fact is useful in forming closed-loop coherent feedback control systems, cf., Section \ref{sec:closed_loop}.

\section{Performance Specifications for Linear Quantum Systems}\label{sec:performance_specification}
In Section \ref{sec:general_sytemes} we have established criteria for stability, passivity, and $L^2$ gain for general quantum dissipative systems studied in Section \ref{sec:open_systems}. These criteria are expressed in terms of operators. In this section we specialize those results to linear quantum systems introduced in Section \ref{sec:models-general}. It can be seen that for linear quantum systems such criteria can be expressed in terms of constant matrices.

\subsection{Stability, Passivity, Gain} \label{sec:stab}

Perhaps the most basic performance characteristic is stability. For system $G$ of open quantum harmonic oscillators presented in Section \ref{sec:models}, stability may be evaluated in terms of the behavior of the number of quanta (e.g. photons) stored in the system, ${\bf N}=a^{\dagger }a=\sum_{j=1}^{n}a_{j}^{\ast }a_{j}$. We introduce the following definition of stability.

\begin{definition}(Stability, \cite[Sec. III-A]{ZJ11})\label{def:stability_2}
Let $w=0$ and $v=0$ in (\ref{eq:gen-1-dyn}), that is there is no energy input to system $G$.  We say that $G$ is (i) \emph{exponentially stable} if there exist scalars $c_0>0, c_1>0$, and $c_2 \geq 0$ such that $\langle {\bf N}(t)\rangle  \leq c_0 e^{-c_1 t } \langle {\bf N} \rangle + c_2$; (ii) \emph{marginally stable} if there exist scalars $c_1>0$ and $c_2 \geq 0$ such that $\langle {\bf N}(t)\rangle \leq c_1 \langle {\bf N} \rangle + c_2 t$; and (iii) \emph{exponentially unstable} if there exists an initial system state and real numbers $c_0>0, c_1>0$ and $c_2$ such that  $\langle {\bf N}(t)\rangle  \geq c_0 e^{c_1 t } \langle {\bf N} \rangle +c_2$.
\end{definition}

For example, for the closed optical cavity in Section \ref{sec:cavity}, Fig.~\ref{fig_closed_cavity}, $a(t) = \exp (-i \omega t) a$, and $a^\ast(t) a(t)=a^\ast a$ for all $t$, which means that $G$ is marginally stable but not exponentially  stable---it oscillates---hence the name \lq\lq{oscillator}\rq\rq. However, an open cavity $(1, \sqrt{\kappa}a, \omega a^\ast a)$  (Fig. \ref{fig_open_cavity}) is exponentially stable, a \emph{damped} oscillator.

The number operator ${\bf N} = a^\dagger a$, whose mean value is the total number of quanta,  is a natural Lyapunov function for $G$, and is directly related to the energy of the system. However we find it more convenient to use storage functions of the form $V=\frac{1}{2} \breve{a}^\dagger P \breve{a}$ for non-negative Hermitian matrices $P$. For such storage functions, the generator function (\ref{eq:generator}) becomes
\begin{equation}\label{eq:generator_linear}
\mathcal{G}_G(V) =\frac{1}{2} \breve{a}^\dag (A^\dag P + PA) \breve{a}.
\end{equation}
With this simple yet important observation, the results in Section \ref{sec:general_sytemes} can be specialized to linear quantum systems.

Define the matrix $F$ by
\begin{equation} \label{F}
Fdt=(d\breve{%
B}^{\#}(t)d\breve{B}^{T}(t))^{T}=
\left[
\begin{array}{cc}
0_{m} & 0  \\
0     &  I_{m}
\end{array}
\right]dt .
\end{equation}

The following result is a simple criterion for stability of linear quantum system $G$, which is a linear version of Theorems \ref{thm:stability} and \ref{thm:strict_passivity_stability}.

\begin{theorem}\label{thm:stability_linear} (Stability, \cite[Theorem 1]{ZJ11})
If there exist constant matrices $P\geq 0$ and $Q\geq cP$ for a scalar $c>0$ such
that
\begin{equation}
A^{\dagger }P+PA+Q\leq 0,
\label{PQ}
\end{equation}%
then inequality
\begin{equation}
\left\langle \breve a^\dagger(t) P \breve a(t)
\right\rangle \leq e^{-ct}\left\langle \breve a^\dagger P \breve a
\right\rangle
+\frac{\lambda}{2c}
\label{V}
\end{equation}%
holds, where $\lambda =\mathrm{tr}[B_{f}^{\dagger }PB_{f}F]$ with $F$ given by (\ref{F}). If also $P \geq  \alpha I$ ($\alpha > 0$), then
$\left\langle a^\dagger(t) a(t)
\right\rangle \leq \frac{1}{\alpha}  e^{-ct}\left\langle \breve a^\dagger P \breve a
\right\rangle
+\frac{\lambda}{2c\alpha} .
$ In this case, $G$ is exponentially stable.
\end{theorem}

In a similar way, by choosing linear versions of supply rate functions positive real lemma and bounded real lemma can be established for linear quantum systems.

In order to simplify the notation we write $u = [w^T \, \, v^T]^T$ for the doubled-up vector of external variables, and define accordingly
\begin{equation}\label{eq:B}
B:=[ B_f \, \, B_d]\left[
\begin{array}{cccc}
I & 0 & 0 & 0 \\
0 & 0 & I & 0 \\
0 & I & 0 & 0 \\
0 & 0 & 0 & I%
\end{array}%
\right],
\end{equation}
where dimensions of identity matrices are implicitly assumed to be conformal to those of $v$ and $w$.

Define a supply rate
\begin{equation}
r(\breve{a}, \breve{u}) = \frac{1}{2}( - \breve{a}^\dagger Q \breve{a} + \breve{u}^\dagger \breve{z} + \breve{z}^\dagger \breve{u}) .
\label{eq:prl-4}
\end{equation}
The we have the following positive real lemma.

\begin{theorem}  \label{thm:PRL} (Positive Real Lemma, \cite[Theorem 3]{ZJ11})
The system $G$ with performance variable $\breve{z} = C_p \breve{a}$  is passive if and only if there exist non-negative definite Hermitian matrices $P$ and $Q$ such that
\begin{equation}
\left[  \begin{array}{cc}
PA + A^\dagger P + Q  & PB -C_p^\dagger
\\
B^\dagger P - C_p  & 0
\end{array} \right ] \leq 0.
\label{eq:prl-111}
\end{equation}
Moreover, $\lambda = \mathrm{tr}[ B_f^\dagger PB_f F] $.
\end{theorem}

\begin{remark}{\rm
When
\[
P = H_0 = \left[
\begin{array}{cc}
\Omega _{-} & \Omega _{+} \\
\Omega _{+}^{\#} & \Omega _{-}^{\#}%
\end{array}%
\right],  V = \frac{1}{2} \breve{a}^\dagger P \breve{a}, \, \, L = C_-a + C_+a^\#, M = K_-a + K_+a^\#,
\]
the operator $Z$ in Eq. (\ref{eq:prl-5}) satisfies $\breve{Z} = \breve{z} =  C_p \breve{a}$. That is, Theorem \ref{thm:PRL} is a special case of Theorem \ref{thm:prl}.
}
\end{remark}

In what follows we discuss $L^2$ gain of linear quantum systems. Denote $D_p = [ D_{pf} \, \, D_{pd}]$, the performance variable can be rewritten as $\breve{z} = C_p \breve{a} + D_p \breve{u}$. Define a supply rate
\begin{equation}
r(\breve{a}, \breve{u}) = - \frac{1}{2} ( \breve{z}^\dagger \breve{z} - g^2 \breve{u}^\dagger \breve{u} ) ,
\label{eq:brl-1}
\end{equation}
where $g \geq 0$ is a real gain parameter. The we have the following results.
\begin{theorem}  \label{thm:BRL} (Bounded  Real Lemma, \cite[Theorem 4]{ZJ11})
The system $G$ with performance variable $\breve{z} = C_p \breve{a}+D_p \breve u$  is bounded real with finite $L^2$ gain less than $g$ if and only if there exists a non-negative Hermitian matrix $P$ such that
\begin{equation}
\left[  \begin{array}{cc}
PA + A^\dagger P +C_p^\dagger C_p   & PB  +C_p^\dagger D_p
\\
B^\dagger  P + D_p^\dagger C_p &  D_p^\dagger D_p-g^2 I
\end{array} \right ] \leq 0.
\label{eq:brl-3}
\end{equation}
 Moreover, $\lambda = \mathrm{tr}[ B_f^\dagger PB_f F] $.
\end{theorem}

\begin{theorem} (Strict Bounded Real Lemma, \cite[Theorem 5]{ZJ11})
\label{thm:brl} The following statements are equivalent.

\begin{description}
\item[i)] The quantum system $G$ defined in (\ref{eq:gen-1-dyn})-(\ref{eq:gen-1-p}) is strictly bounded real with disturbance attenuation $g$.

\item[ii)] $A$ is stable and $\left\Vert C_{p}\left( sI-A\right)^{-1}B+D_{p}\right\Vert _{\infty }<g$.

\item[iii)] $g^{2}I-D_{p}^{\dagger }D_{p}>0$ and there exists a Hermitian matrix $P_{1}>0$ satisfying inequality%
\begin{equation}
\left[
\begin{array}{ccc}
A^{\dagger }P_{1}+P_{1}A & P_{1}B & C_{p}^{\dagger } \\
B^{\dagger }P_{1} & -gI & D_{p}^{\dagger } \\
C_{p} & D_{p} & -gI%
\end{array}%
\right] <0.  \label{LMI_ori}
\end{equation}

\item[iv)] $g^{2}I-D_{p}^{\dagger }D_{p}>0$ and there exists a Hermitian matrix $P_{2}>0$ satisfying the algebraic Riccati equation
\begin{eqnarray*}
&&A^{\dagger }P_{2}+P_{2}A +\left( P_{2}B+C_{p}^{\dagger
}D_{p}\right)   \\
 && \times\left( g^{2}I-D_{p}^{\dagger }D_{p}\right) ^{-1}(
B^{\dagger }P_{2}^{\dagger }+D_{p}^{\dagger }C_{p})  \\
&=&0
\end{eqnarray*}
with $A+BB^{\dagger }P_{2}$ being Hurwitz.
\end{description}

Furthermore, if these statements hold, then $P_1 < P_2$.
\end{theorem}

\subsection{LQG Performance} \label{sec:analysis-lqg}

In this section a quantum LQG cost function is first defined in the annihilation-creation form, and whose evaluation is connected to a Lyapunov equation in the complex domain. After that the real domain case is presented. More discussions can be found in, e.g., \cite{NJP09,ZJ11}.

Consider the following stable linear quantum system
\begin{equation}
d\breve{a}(t)=A\breve{a}(t)dt+B_{f}d\breve{B}(t)  \label{system_LQG}
\end{equation}%
where $B(t)$ is a quantum Wiener process introduced in Section \ref{sec:field}. Given a performance variable  $\breve{z}(t)=C_{p}\breve{a}(t)$, along the line of \cite{NJP09},  the infinite-horizon LQG cost is
\begin{eqnarray}
\mathfrak{J}_{\infty} &:=&\lim_{t_{f}\rightarrow \infty }\frac{1}{t_{f}} \int_{0}^{t_{f}}\frac{1}{2}\left\langle \breve{z}^\dag (t)%
\breve{z}(t)+\breve{z}^{T}(t)\breve{z}^{\#}(t)\right\rangle dt \label{LQG_index}
\\
&=&\lim_{t_{f}\rightarrow \infty }\frac{1}{t_{f}}\int_{0}^{t_{f}}\mbox{Tr}\left\{C_{p}P_{LQG}(t)C_{p}^\dag \right\} dt  \nonumber
\\
&=&\mbox{Tr}\left\{C_{p}P_{LQG}C_{p}^\dag \right\} , \nonumber
\end{eqnarray}
where the constant Hermitian matrix $P_{LQG} \geq 0$ satisfies the following Lyapunov equation%
\begin{equation}
AP_{LQG}+P_{LQG}A^\dag +\frac{1}{2}B_{f}B_{f}^\dag =0.
\label{lyap}
\end{equation}

In quadrature form, given a stable linear quantum system
\begin{equation}
d\tilde{a}(t)=\tilde{A}\tilde{a}(t)dt+\tilde{B}_{f}d\tilde{B}(t)  \label{system_LQG2}
\end{equation}%
with performance variable  $\tilde{z}(t)=\tilde{C}_{p}\tilde{a}(t)$. Assume that the constant real matrix $\tilde{P}_{LQG}\geq 0$ is the (unique) solution to  the following Lyapunov equation in the real domain
\begin{equation}
\tilde{A}\tilde{P}_{LQG}+\tilde{P}_{LQG}\tilde{A}^T+\tilde{B}_{f}\tilde{B}_{f}^T=0.
\label{lyap_2}
\end{equation}
Then
\begin{equation}
\mathfrak{J}_{\infty} = \mbox{Tr}\left\{\tilde{C}_{p}\tilde{P}_{LQG}\tilde{C}_{p}^\dag \right\}.
\label{LQG_index2}
\end{equation}

\section{Coherent Feedback Control}\label{sec:feedback}
We have discussed interconnections of quantum systems (Section \ref{sec:interconnection}), open linear quantum systems (Section \ref{sec:models-general}), and their performance specifications (Section \ref{sec:performance_specification}). We are now in a position to study synthesis of open linear quantum systems; that is, how to connect a plant of interest to another system (namely controller) so as to achieve pre-specified control performance.

\subsection{Closed-Loop Plant-Controller System}\label{sec:closed_loop}
\begin{figure}
\centering
  \includegraphics[width=2.5in]{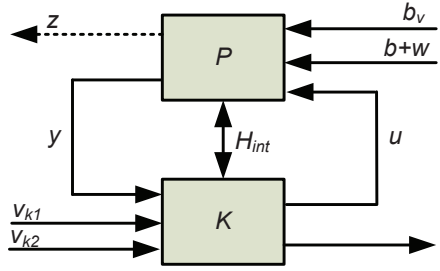}\\
  \caption{Coherent feedback control arrangement}\label{closed-loop}
\end{figure}

In Figure \ref{closed-loop}, $P$ is the plant to be controlled, $K$ is the controller to be designed. Clearly, this feedback system involves both direct and indirect couplings between $P$ and $K$.

The plant $P$  is described by a system of quantum stochastic differential equations (QSDEs)
\begin{eqnarray}
\dot{\breve{a}}(t) &=&A\breve{a}(t)+B_{12} \breve{a}_K(t)+
B_{v}\breve{b}_{v}(t)+B_{f}\breve{w}(t) \nonumber  \\
&&+B_{f}\breve{b}(t)+B_{u}\breve{u}(t), \, \, \breve{a}(0)=\breve{a},  \nonumber \\
\breve{y}(t) &=&C\breve{a}(t)+D_{v}\breve{b}_{v}(t)+D_{f}\breve{w}%
(t)+D_{f}\breve{b}(t).  \label{plant_optical}
\end{eqnarray}%
The inputs $\breve{w}(t)$ and $\breve{b}(t)$ are defined in Section \ref{sec:models-general}. $\breve{y}(t)$ is a selection of output field channels from the plant. $\breve{b}_{v}(t)$ is a vector of additional quantum white noises; $\breve{u}(t)$ is a quantum field signal from the to-be-designed controller $K$, hence it is a vector of physical variables.  The term $B_{12} \breve{a}_K(t)$ is due to direct coupling between $P$ and $K$.

The fully quantum controller $K$ is a linear quantum system of the form\footnote{We assume that all the variables and matrices of the plant and the controller have compatible dimension, but we don't bother to specify them explicitly.}
\begin{eqnarray}
\dot{\breve{a}}_{K}(t) &=&A_{K}\breve{a}_{K}(t)+B_{21}\breve{a}(t)+B_{K}\breve{y}(t)+B_{K1}\breve{b}%
_{v_{K1}}(t) \nonumber \\
&&+B_{K2}\breve{b}%
_{v_{K2}}(t), \, \, \breve{a}_{K}(0)=\breve{a}_K,   \nonumber \\
\breve{u}(t) &=&C_{K}\breve{a}_{K}(t)+\breve{b}_{v_{K1}}(t).   \label{controller_optical2}
\end{eqnarray}%
This structure allows for direct coupling and indirect coupling between the plant $P$ and the controller $K$. Here, $\breve{b}_{v_{K1}}(t)$ and $\breve{b}_{v_{K2}}(t)$ are independent quantum white noises, and $\breve{u}(t)$ is the field output of the controller  corresponding to $\breve{b}_{v_{K1}}(t)$. Finally the terms $B_{12} \breve{a}_K(t)$ and $B_{21}\breve{a}(t)$ are due to the direct coupling between the plant and controller in terms of an interaction Hamiltonian
\begin{equation}
\label{syn:Hamiltonian}
H_{int}= \frac{1}{2}\left( \breve{a}^\dag \Xi^\dag \breve{a}_{K}+%
\breve{a}_{K}^\dag \Xi\breve{a}\right),
\end{equation}
where
$
\Xi=\Delta (iK_{-},iK_{+})
$
for complex matrices $K_{-}$ and $K_{+}$ of suitable dimensions, cf. Section \ref{sec:models-direct}.

The controller matrices $K_-, K_+$,  (or $B_{12}, B_{21}$) for direct coupling, and $A_K, B_K, C_K, B_{K1}, B_{K2}$ for indirect coupling are to be found to optimize performance criteria defined in terms of the  closed-loop performance variable
\begin{equation}
\breve{z}(t) =[ C_{p} ~~ D_{u}C_K ]\left[
\begin{array}{c}
\breve{a}(t) \\
\breve{a}_{K}(t) \\
\end{array}%
\right] +  \breve{D}_{pf}\breve{w}(t) .
\label{syn_performance}
\end{equation}%

Because standard matrix algorithms will be used in $H^\infty$ synthesis and LQG synthesis in later sections, we resort to quadrature representation discussed in Section \ref{sec:models-quadrature}. Let $\tilde{a}$, $\tilde{a}_{K}$, $\tilde{w}$, $\tilde{b}$, $\tilde{b}_{v}$, $\tilde{u}$, $\tilde{z}$, $\tilde{y}$, $\tilde{b}_{v_{K1}}$,  $\tilde{b}_{v_{K2}}$  be the quadrature counterparts of $\breve{a}$, $\breve{a}_{K}$, $\breve{w}$, $\breve{b}$, $\breve{b}_{v}$, $\breve{z}$, $\breve{\beta}_{u}$, $\breve{y}$, $\breve{b}_{v_{K1}}$, $\breve{b}_{v_{K2}}$ respectively. Define
$$
\tilde{A}_{cl} =\left[
\begin{array}{cc}
\tilde{A} & \tilde{B}_{u}\tilde{C}_{K} \\
\tilde{B}_{K}\tilde{C} & \tilde{A}_{K}%
\end{array}%
\right]+\tilde{\Xi} , ~~\ \tilde{B}_{cl} = \left[
\begin{array}{c}
\tilde{B}_{f} \\
\tilde{B}_{K}\tilde{D}_{f}%
\end{array}%
\right] ,
$$
$$ \tilde{G}_{cl}=\left[
\begin{array}{cccc}
\tilde{B}_{f} & \tilde{B}_{v} & \tilde{B}_{u} & 0 \\
\tilde{B}_{K}\tilde{D}_{f} & \tilde{B}_{K}\tilde{D}_{v} & \tilde{%
B}_{K1} & \tilde{B}_{K2}%
\end{array}%
\right] ,
$$
$$
\tilde{C}_{cl} =\left[
\begin{array}{cc}
\tilde{C}_{p} & \tilde{D}_{u}\tilde{C}_{K}%
\end{array}%
\right] ,~~\tilde{D}_{cl} = \tilde{D}_{pf},
$$
where $\tilde{\Xi} =[0 ~ \tilde{B}_{12}; \tilde{B}_{21} ~ 0]$ satisfies $\tilde{B}_{21} = \Theta\tilde{B}_{12}^{T}\Theta$. Then the closed-loop system in the quadrature representation is given by%
\begin{eqnarray}
\left[
\begin{array}{c}
\dot{\tilde{a}}(t) \\
\dot{\tilde{a}}_{K}(t)%
\end{array}%
\right]  &=& \tilde{A}_{cl}\left[
\begin{array}{c}
\tilde{a}(t) \\
\tilde{a}_{K}(t)%
\end{array}%
\right] +\tilde{B}_{cl}\tilde{w}(t)+\tilde{G}_{cl}\left[
\begin{array}{c}
\tilde{b}(t) \\
\tilde{b}_{v}(t) \\
\tilde{b}_{v_{K1}}(t) \\
\tilde{b}_{v_{K2}}(t)%
\end{array}%
\right] , \label{system_a} \\
\tilde{z}(t) &=& \tilde{C}_{cl}\left[
\begin{array}{c}
\tilde{a}(t) \\
\tilde{a}_{K}(t)%
\end{array}%
\right]+\tilde{D}_{cl}\tilde{w}(t) . \label{system_b}
\end{eqnarray}

\subsection{$H^{\infty}$ Control} \label{sec:h-infty-synthesis}
As in the classical case, the bounded real lemmas stated in Section \ref{sec:stab} can be used for $H^\infty$ controller synthesis of open linear quantum systems. It is shown in \cite{JNP08} that for open linear quantum systems $H^\infty$ control performance and physical realizability condition of controllers can be treated separately. Adding direct coupling between plants and controllers complicates $H^\infty$ controller synthesis. Nonetheless, the separation of $H^\infty$ control performance and physical realizability condition still holds. This is a unique feature of quantum $H^\infty$ controller synthesis: To guarantee physical realizability, vacuum noise is added, while such noise does not affect $H^\infty$ control performance \cite{JNP08}.

\subsubsection{LMI Formulation}

In this section we present a general formulation using LMIs for $H^\infty$ synthesis of open linear quantum stochastic systems.

According to the strict bounded real lemma (Theorem \ref{thm:brl}), the closed-loop system (\ref{system_a})-(\ref{system_b}) is internally stable and strictly bounded real (from $\tilde{w}$ to $\tilde{z}$) with disturbance attenuation $g$ if and only if there is a real symmetric matrix $\mathcal{P}$ such that%
\begin{eqnarray}
\mathcal{P} & > & 0  \label{LMI} \\
\left[
\begin{array}{ccc}
\tilde{A}_{cl}^{T}\mathcal{P}+\mathcal{P}\tilde{A}_{cl} & \mathcal{P}\tilde{B}_{cl} &
\tilde{C}_{cl}^{T} \\
\tilde{B}_{cl}^{T}\mathcal{P} & -gI & \tilde{D}_{cl}^{T} \\
\tilde{C}_{cl} & \tilde{D}_{cl} & -gI%
\end{array}%
\right] &<&0.  \label{LMI2}
\end{eqnarray}%

The $H^\infty$ controller synthesis is to find indirect coupling parameters $\tilde{A}_{K}$, $\tilde{B}_{K}$, $\tilde{C}_{K}$ and direct coupling parameters $\tilde{\Xi} $ such that Eqs. (\ref{LMI})-(\ref{LMI2}) hold.

Partition $\mathcal{P}$ and its inverse $\mathcal{P}^{-1}$ to be
$$
\mathcal{P=}\left[
\begin{array}{cc}
\mathbf{Y} & N \\
N^{T} & \ast%
\end{array}%
\right] , ~~ \mathcal{P}^{-1}=\left[
\begin{array}{cc}
\mathbf{X} & M \\
M^{T} & \ast%
\end{array}%
\right] .
$$%
Define matrices
$$
\Pi _{1}=\left[
\begin{array}{cc}
\mathbf{X} & I \\
M^{T} & 0%
\end{array}%
\right] , ~~ \Pi _{2}=\left[
\begin{array}{cc}
I & \mathbf{Y} \\
0 & N^{T}%
\end{array}%
\right] .
$$%
And also define a change of variables%
\begin{eqnarray}
\mathbf{\hat{A}} &\mathbf{=}&N( \tilde{A}_{K}M^{T}+\tilde{B}_{K}%
\tilde{C}\mathbf{X}) +\mathbf{Y}( \tilde{B}_{u}\tilde{C}%
_{K}M^{T}+\tilde{A}\mathbf{X}) ,  \nonumber \\
\mathbf{\hat{B}} &\mathbf{=}&N\tilde{B}_{K},  \nonumber \\
\mathbf{\hat{C}} &\mathbf{=}&\tilde{C}_{K}M^{T},  \nonumber \\
\mathbf{\Omega } &=&\Pi _{1}^{T}\mathcal{P}\tilde{\Xi} \Pi _{1}.   \label{congruence}
\end{eqnarray}%
With these notations, (\ref{LMI}) and  (\ref{LMI2}) hold if and only if the following inequalities hold.%
\begin{equation}
-\left[
\begin{array}{cc}
X & I \\
I & Y%
\end{array}%
\right] <0 ,  \label{LMIs2}
\end{equation}%
\begin{eqnarray}
&& \hspace{-6mm} \left[
\begin{array}{c}
\tilde{A}\mathbf{X+X}\tilde{A}^{T}+\tilde{B}_{u}\mathbf{\hat{C}+}(\tilde{B}%
_{u}\mathbf{\hat{C}})^{T} \\
\mathbf{\hat{A}+}\tilde{A}^{T} \\
\tilde{B}_{f}^{T} \\
\tilde{C}_{p}\mathbf{X+}\tilde{D}_{u}\mathbf{\hat{C}}%
\end{array}%
\right.   \nonumber \\
&&\hspace{1cm} \left.
\begin{array}{ccc}
\tilde{A}+\mathbf{\hat{A}}^{T} & \ast  & \ast  \\
\tilde{A}^{T}\mathbf{Y}+\mathbf{Y}\tilde{A}+\mathbf{\hat{B}}\tilde{C}+(%
\mathbf{\hat{B}}\tilde{C})^{T} & \ast  & \ast  \\
(\mathbf{Y}\tilde{B}_{f}+\mathbf{\hat{B}}\tilde{D}_{f})^{T} & -gI & \ast  \\
\tilde{C}_{p} & \tilde{D}_{cl} & -gI%
\end{array}%
\right]  \nonumber \\
&&\hspace{-8mm}+\left[
\begin{array}{cc}
\tilde{B}_{12}M^{T}+(\tilde{B}_{12}M^{T})^{T} & (N\tilde{B}_{21}\mathbf{X}%
)^{T}+(\mathbf{Y}\tilde{B}_{12}M^{T})^{T} \\
N\tilde{B}_{21}\mathbf{X}+\mathbf{Y}\tilde{B}_{12}M^{T} & N\tilde{B}_{21}+(N%
\tilde{B}_{21})^{T} \\
0 & 0 \\
0 & 0%
\end{array}%
\right.  \nonumber  \\
&& \hspace{6.5cm} \left.
\begin{array}{cc}
0 & 0 \\
0 & 0 \\
0 & 0 \\
0 & 0%
\end{array}%
\right]  \nonumber  \\
&<&0. \label{LMIs}
\end{eqnarray}
If (\ref{LMIs2}) and (\ref{LMIs}) are simultaneously soluble, according to Eq. (\ref{congruence}), the following matrices can be obtained.
\begin{eqnarray}
\tilde{B}_{K} & \hspace{-2mm}  =& \hspace{-2mm} N^{-1}\mathbf{\hat{B}},  \label{solution} \\
\tilde{C}_{K} & \hspace{-2mm}  =& \hspace{-2mm} \mathbf{\hat{C}}\left( M^{T}\right) ^{-1},  \nonumber \\
\tilde{A}_{K} & \hspace{-2mm}  =& \hspace{-2mm} N^{-1}( \mathbf{\hat{A}-}N\tilde{B}_{K}\tilde{C}\mathbf{X-Y}( \tilde{B}_{u}\tilde{C}_{K}M^{T}+\tilde{A}\mathbf{X}))M^{-T},  \nonumber \\
\tilde{\Xi} & \hspace{-2mm}  =& \hspace{-2mm} \mathcal{P}^{-1}\left( \Pi _{1}^{-T}\right)\mathbf{\Omega }\Pi
_{1}^{-1}.  \label{interaction}
\end{eqnarray}%
Unfortunately, notice that there are such terms as $N\tilde{B}_{21}\mathbf{X}$ and $\mathbf{Y}\tilde{B}_{12}M^{T}$ in inequality (\ref{LMIs}), which induce nonlinearity. The above analysis shows it is hard to directly utilize LMI techniques to do controller design when direct coupling is involved.

\subsubsection{Multi-step Optimization} \label{algorithm}

In this section, we attempt to circumvent the above difficulty by proposing a multi-step optimization procedure which is formulated as follows:

\textit{Initialization}. Set $\tilde{B}_{12 }=0$ and $\tilde{B}_{21 }=0$.

\textit{Step 1}. Solve linear matrix inequalities (\ref{LMIs2}) and (\ref{LMIs}) for parameters $\mathbf{\hat{A}}$, $\mathbf{\hat{B}}$, $\mathbf{\hat{C}}$, $\mathbf{X}$, $\mathbf{Y}$ and disturbance gain $g$, then choose matrices $M$ and $N$ satisfying $MN^{-1}=I-XY$.

\textit{Step 2. } Pertaining to \textit{Step 1}. Solve inequality (\ref{LMIs}) for direct coupling parameters $\tilde{B}_{12}, \tilde{B}_{21}$  and disturbance gain $g$.

\textit{Step 3. } Fix $\tilde{B}_{12 }$ and $\tilde{B}_{21 }$ obtained in \textit{Step 2} and $M $ and $N$ in \textit{Step 1}, go to \textit{Step 1}.

After the above iterative procedure is complete, use the values  $\tilde{B}_{12}, \tilde{B}_{21}$, $\tilde{A}_K, \tilde{B}_K, \tilde{C}_K$ obtained to find $\tilde{B}_{K1}, \tilde{B}_{K2}$ to ensure physical realizability of the controller. A complete procedure of finding matrices $\tilde{B}_{K1}, \tilde{B}_{K2}$ is given in \cite[Sec. V-D]{JNP08}.

\begin{remark}
{\rm \textit{Steps 1 and 2 } are standard LMI problems which can be solved efficiently using the Matlab LMI toolbox. However, there is some delicate issue in \textit{Step 3.} Assume that $\tilde{B}_{12 }$ and $\tilde{B}_{21 }$ have been obtained in \textit{Step 2}. According to the second item in (\ref{LMIs}), constant matrices $M$ and $N$ must be specified in order to render (\ref{LMIs}) linear in parameters $\mathbf{\hat{A}}$, $\mathbf{\hat{B}}$, $\mathbf{\hat{C}}$, $\mathbf{X}$, $\mathbf{Y}$, and disturbance gain $g$. In \textit{Step 3}, $M$ and $N$ obtained in \textit{Step 1 } is used. Unfortunately, this choice of $M$ and $N$ sometimes may generate a controller whose parameters are ill-conditioned. Due to this reason, $M$ and $N$ in \textit{Step 3} might have to be chosen carefully to produce a physically meaningful controller. This fact is illuminated by an example in \cite[Sec. IV-C6]{ZJ11}.
}
\end{remark}

Finally we discuss robustness briefly. It is demonstrated in \cite{ZJ11} that direct coupling may improve robustness of closed-loop quantum feedback systems. For instance, for the example studied in \cite[Sec. VII.A]{JNP08}, using coupling coefficients $\kappa_1 = 2.6$, $\kappa_2 = \kappa_3 = 0.2$, we implement Step 1 of the above multi-step optimization procedure to design an indirect coupling, and obtain closed-loop $L^2$ gain 0.0487.  We implement Step 2 to design direct coupling and obtain an $L^2$ gain of 0.0498. This is a bit worse than the previous one, however the difference is quite small. Now we assume there is uncertainty in the coupling coefficient $\kappa_1$, say the actual value of $\kappa_1$ is 1.3. In this case, the $L^2$ gain of the closed-loop with indirect coupling becomes  0.1702, which is a significant performance degradation. However, the $L^2$ gain of the closed-loop with both direct and indirect couplings is 0.0595, which is still close to the original 0.0498.

\subsection{LQG Control} \label{sec:LQG-synthesis}

In this section we study the problem of coherent quantum LQG control by means of both direct and indirect couplings. In contrast to the coherent quantum $H^\infty$ controller synthesis presented in Section \ref{sec:h-infty-synthesis}, the nice property of separation of control and physical realizability does not hold any more. This is evident as LQG control concerns the influence of quantum white noise on the plant, the addition of quantum noise that guarantees the physical realizability of the to-be-designed controller affects the overall LQG control performance.

In the following we just give a brief formulation of the coherent quantum LQG control problem. In-depth discussions can be found in \cite{NJP09,ZJ11, VP11a,VP11b}.

We make the following assumption.

\noindent \emph{Assumption A2.} There are no quantum signal $\breve{w}(t)$ and noise input $\breve{b}_{v}(t)$ in the quantum plant $P$ in (\ref{plant_optical}).

Following the development in Section \ref{sec:analysis-lqg}, the LQG control objective is to design a controller (\ref{controller_optical2}) such that the performance index $\mathfrak{J}_{\infty}=\mbox{Tr}\left\{ \tilde{C}_{cl}\tilde{P}_{LQG}\tilde{C}_{cl}^\dag \right\} $ is minimized, subject to equation (\ref{lyap_2}) and the quadrature counterpart of the physical realizability condition (\ref{eq:c-2-ccr})-(\ref{eq:c-2-d}).

As yet, quantum LQG coherent feedback is still an outstanding problem, there are no analytic solutions.  In \cite{NJP09} an indirect coupling is designed to address the coherent quantum LQG control problem, where a numerical procedure based on semidefinite programming is proposed to design the indirect coupling. In order to design both direct and indirect couplings. In \cite[Sec. IV-D]{ZJ11}  a multi-step optimization algorithm is developed to incorporate direct coupling into numerical design procedures.  Some theoretical insights into the structure of quantum LQG coherent feedback control have been provided in \cite{VP11a,VP11b}.

\section{Network Synthesis} \label{sec:synthesis}

A linear quantum controller, obtained from either coherent $H^\infty$ or LQG control synthesis, is in the form of a set of linear quantum stochastic differential equations. Network synthesis theory is concerned with how to physically implement such controllers by means of physical devices like optical instruments.  This problem has been addressed in \cite{NJD09,Nur10a,Nur10b,petersen11}. The general result is: A general linear quantum dynamical system can be (approximately) physically implemented by linear and nonlinear quantum optical elements such as optical cavities, parametric oscillators, beam splitters, and phase shifters.

Lately, the Mabuchi group at Stanford \cite{TNP+11} has developed a Quantum Hardware Description Language (QHDL) to facilitate the analysis and synthesis of quantum feedback networks described in this survey. As a subset of the standard Very High Speed Integrated Circuit (VHSIC) Hardware Description Language (VHDL), QHDL provides high-level modular representations of quantum feedback networks.This user- friendly interface will be helpful to the design of complex photonic circuit models and their analysis and simulation.

\section{Conclusions} \label{sec:conclusion}

In this survey we have presented a brief look at recent results concerning quantum feedback networks and control. On the basis of this model interconnection structures of quantum systems have been presented. Fundamental characteristics of quantum systems such as stability, passivity, and $L^2$ gain have been described. It turns out that for linear quantum systems these fundamental characteristics have very explicit forms. The problem of coherent $H^\infty$ control and coherent LQG control have been discussed.

\section*{Acknowledgement}
The first author wishes to thank Daoyi Dong and Hu Zhang for their helpful comments.


\begin{thebibliography}{99}
\bibitem{AASDM02}
Armen M A, Au J K,  Stockton J K, Doherty A C, Mabuchi H.  Adaptive homodyne measurement of optical phase. Physical Review Letters, 2002, 89:133602


\bibitem{Belavkin83}
Belavkin V P. On the theory of controlling observable quantum systems. Automation and Remote Control, 1983, 44(2):178-188

\bibitem{BBC+93}
Bennett C H, Brassard G, Crepeau C, Jozsa R, Peres A, Wootters W K. Teleporting an unknown quantum state via dual classical and Einstein-Podolsky-Rosen channels. Physical Review Letters, 1993, 70:1895-1899

\bibitem{BvHJ07}
Bouten L, van Handel R, James M R. An introduction to quantum filtering. SIAM J. Contr. Optimiz., 2007, 46(6):2199-2241


\bibitem{BCR10}
Brif C, Chakrabarti R, Rabitz H. Control of quantum phenomena: past, present and future. New J. Phys., 2010, 12:075008

\bibitem{Car93}
Carmichael H J. Quantum trajectory theory for cascaded open systems. Phys. Rev. Lett., 1993, 70:2273

\bibitem{DJ99}
Doherty A C, Jacobs K. Feedback-control of quantum systems using continuous state-estimation. Physical Review  A, 1999, 60:2700-2711


\bibitem{DHJMT00}
Doherty A C, Habib S, Jacobs K, Mabuchi H, and Tan S M. Quantum feedback control and classical control theory. Physical Review  A, 2000, 62:012105

\bibitem{DP10}
Dong D, Petersen I R. Quantum control theory and applications: a survey. IET Control Theory \& Applications, 2010, 4(12):2651-2671


\bibitem{DM03}
Dowling J P, Milburn G J. Quantum technology: the second quantum revolution. Philosophical Transactions of the Royal Society A: Mathematical, Physical and Engineering Sciences, 2003, 361(1809):1655-1674


\bibitem{EB05}
Edwards S C, Belavkin V P.  Optimal quantum filtering and quantum feedback control, arXiv:quant-ph/0506018v2 [quant-ph], 2005

\bibitem{Gar93}
Gardiner C W. Driving a quantum system with the output field from another driven quantum system. Phys. Rev. Lett., 1993, 70:2269

\bibitem{GZ04}
Gardiner G,  Zoller P. Quantum Noise. Springer, Berlin, 2004.

\bibitem{Gough08}
Gough J E. Construction of bilinear control Hamiltonians using the series product and quantum feedback. Phys. Rev. A, 2008, 78:052311

\bibitem{GJ09}
Gough J E, James M R. Quantum feedback networks:  Hamiltonian formulation. Commun. Math. Phys., 2009, 287:1109-1132


\bibitem{GJ09b}
Gough J E, James M R. The series product and its application to quantum feedforward and feedback networks. IEEE Transactions on Automatic Control, 2009, 54(11):2530-2544

\bibitem{GW09}
Gough J E, Wildfeuer S. Enhancement of field squeezing using coherent feedback. Phys. Rev. A, 2009, 80:042107

\bibitem{GJN10}
Gough J E, James M R, Nurdin H I. Squeezing components in linear quantum feedback networks. Phys. Rev. A, 2010, 81:023804


\bibitem{HP10}
Harno H, Petersen I R. Coherent control of linear quantum systems: a differential evolution approach. Proc. American Control Conference, 2010, 1912-1917


\bibitem{HP84}
Hudson R L, Parthasarathy  K R. Quantum Ito's formula and stochastic evolutions. Commun. Math. Phys, 1984, 93:301¨C323



\bibitem{IYY+11}
Iida S, Yukawa M, Yonezawa H, Yamamoto N, Furusawa A. Experimental demonstration of coherent feedback control on optical field squeezing, arXiv:1103.1324v1 [quant-ph], 2011


\bibitem{JNP08}
James M R, Nurdin H I, Petersen I R. $H^{\infty }$ control of linear quantum stochastic systems. IEEE Trans. Automat. Control, 2008, 53(8):1787-1803

\bibitem{JG10}
James M R, Gough J E. Quantum dissipative systems and feedback control design by interconnection. IEEE Trans. Automat. Control, 2010, 55(8):1806-1821


\bibitem{JK10}
James M R, Kosut R L. Quantum Estimation and Control. In The Control Handbook, Second Edition, Edited by William S. Levine, CRC Press, 2010:31-1-31-42


\bibitem{KNP+10}
Kerckhoff J, Nurdin H I, Pavlichin D S,  Mabuchi H. Coherent-feedback formulation of a continuous quantum error correction protocol. Physical Review Letters, 2010, 105:040502

\bibitem{KPC+11}
Kerckhoff J, Pavlichin D S, Chalabi H, Mabuchi H. Design of nanophotonic circuits for autonomous subsystem quantum error correction. New Journal of Physics, 2011, 13(5):055022


\bibitem{Lloyd00}
Lloyd S. Coherent quantum feedback. Phys. Rev. A, 2000, 62:022108


\bibitem{MP11}
Maalouf A, Petersen I R. Coherent $H_{\infty}$ control for a class of linear complex quantum systems. IEEE Trans. Automat. Contr., 2011, 56(2):309-319

\bibitem{MP11a}
Maalouf A, Petersen I R. Bounded real properties for a class of linear complex quantum systems. IEEE Trans. Automat. Contr., 2011, 56(4):786-801


\bibitem{MK05}
Mabuchi H, Khaneja N. Principles and applications of control in quantum systems. Int. J. Robust Nonlinear Control, 2005, 15:647-667


\bibitem{Mabuchi08}
Mabuchi M. Coherent-feedback quantum control with a dynamic compensator. Phys. Rev. A, 2008, 78:032323


\bibitem{Mabuchi11}
Mabuchi M. Coherent-feedback control strategy to suppress spontaneous switching in ultralow power optical bistability. Appl. Phys. Lett., 2011, 98:193109

\bibitem{MPS+09}
Matthews J C F, Politi A, Stefanov A, O'Brien J L. Manipulation of multiphoton entanglement in waveguide quantum circuits. Nature Photonics 2009, 3:346-350

\bibitem{MAK+97}
Mewes  M O, Andrews M R, Kurn D M, Durfee D S, Townsend C G,  Ketterle W. Output coupler for Bose-Einstein condensed atoms. Physical Review Letters, 1997, 78:582-585

\bibitem{MvH07}
Mirrahimi M, van Handel R. Stabilizing feedback controls for quantum system. SIAM Journal on Control and Optimization, 2007, 46(2):445-467


\bibitem{NWC+00}
Nelson R J, Weinstein Y, Cory D, Lloyd S. Experimental demonstration of fully coherent quantum feedback. Phys. Rev. Lett., 2000, 85(14):3045-304


\bibitem{NJP09}
Nurdin H I, James M R, Petersen I R. Coherent quantum LQG control. Automatica, 2009, 45:1837-1846

\bibitem{NJD09}
Nurdin H I, James M R, Doherty A C. Network synthesis of linear dynamical quantum stochastic systems. SIAM J. Control and Optim., 2009, 48(4):2686-2718

\bibitem{Nur10a}
Nurdin H I. Synthesis of linear quantum stochastic systems via quantum feedback networks. IEEE Trans. Automat. Control, 2010, 55(4):1008-1013

\bibitem{Nur10b}
Nurdin H I. On synthesis of linear quantum stochastic systems by pure cascading. IEEE Trans. Automat. Control, 2010, 55(10):2439-2444

\bibitem{Pa92}
Parthasarathy K R. An Introduction to Quantum Stochastic Calculus. Berlin, Germany: Birkhauser, 1992.

\bibitem{petersen11}
Petersen I R. Cascade cavity realization for a class of complex transfer functions arising in coherent quantum feedback control. Automatica, 2011, 47(8):1757-1763

\bibitem{petersen12}
Petersen I R. Low frequency approximation for a class of linear quantum systems using cascade cavity realization. Systems \& Control Letters, 2012, 61(1):173-179

\bibitem{PMO+09}
Politi A,  Matthews J C F, O'Brien J L. Shor's quantum factoring algorithm on a photonic chip. Science, 2009, 325(5945):1221

\bibitem{SSV+10}
Sansoni L, Sciarrino F, Vallone G, Mataloni P, Crespi A, Ramponi R, Osellame  R. Polarization Entangled State Measurement on a Chip. Phys. Rev. Lett., 2010, 105(4):200503

\bibitem{SSM08}
Sarma G, Silberfarb A, Mabuchi H. Quantum stochastic calculus approach to modeling double-pass atom-field coupling. Phys. Rev. A, 2008, 78:025801


\bibitem{SP09}
Shaiju A J, Petersen I R. On the physical realizability of general linear quantum stochastic differential dquations with complex coefficients. Proc. 48h IEEE Conference on Decision and Control (CDC), 2009, 1422-1427

\bibitem{SM06}
Sherson J F, Molmer K. Polarization squeezing by optical Faraday rotation. Phys. Rev. Lett., 2006, 97:143602

\bibitem{Sho94}
Shor P W. Algorithms for quantum computation: discrete logarithms and factoring.
In: Proceedings of the 35th Annual Symposium on Foundations of Computer Science (IEEE Press, Los Alamitos, CA), 1994, 124-134

\bibitem{SKT+09}
Smith B J, Kundys D, Thomas-Peter N,  Smith P G R,  Walmsley I A. Phase-controlled integrated photonic quantum circuits. Opt. Express, 2009, 17(16):13516-13525

\bibitem{TNP+11}
Tezak N, Niederberger A, Pavlichin D S, Sarma G, Mabuchi H. Specification of photonic circuits using quantum hardware description language.  arXiv:1111.3081v1 [quant-ph], 2011

\bibitem{vdS96}
 van der Schaft A. $L_2$ Gain and Passivity Techniques in Nonlinear Control. Springer-Verlag New York, Inc. Secaucus, NJ, USA, 1996.

\bibitem{vHSM05}
van Handel R, Stockton J K, and Mabuchi H. Feedback control of quantum state reduction. IEEE Transactions on Automatic Control, 2005, 50(6):768-780


\bibitem{WM08}
Walls D F and Milburn G J. Quantum Optics, 2nd edition, 2008, Springer


\bibitem{WNZ+11}
Wang S, Nurdin H I, Zhang G, James M R. Implementation of classical linear stochastic systems using quantum optical components. Proc. 2011 Australian Control Conference, pp. 352-357, Melbourne, Australia,.


\bibitem{Wil72}
Willems J C. Dissipative dynamical systems - Part I: general theory. Archive Rat. Mech. Anal., 1972, 45:321-351

\bibitem{WT02}
Willems J C, and Trentelman H L. Synthesis of dissipative systems using quadratic differential forms: part I. IEEE Trans. Automatic Control, 2002, 47(1):53-69


\bibitem{Wil07}
Willems J C. The behavioral approach to open and interconnected systems. IEEE Control Systems Magazine, 2007, 27(6):46-99

\bibitem{WM93}
Wiseman H M, Milburn G J. Quantum theory of optical feedback via homodyne detection. Physical Review Letters,  1993, 70:548-551

\bibitem{WM94b}
Wiseman H M, Milburn G J. All-optical versus electro-optical quantum-limited feedback. Phys. Rev. A, 1994, 49:4110-4125


\bibitem{WM10}
Wiseman H M, Milburn G J. Quantum measurement and control. Cambridge University Press, New York, 2009.


\bibitem{YK03a}
Yanagisawa M, Kimura H. Transfer function approach to quantum control-part I: dynamics of quantum feedback systems.
IEEE Trans. Automatic Control, 2003, 48:2107-2120


\bibitem{YK03b}
Yanagisawa M, Kimura H. Transfer function approach to quantum control-part II: control concepts and applications.
IEEE Trans. Automatic Control, 2003, 48:2121-2132

\bibitem{VP11a}
Vladimirov I.G, Petersen I R. A quasi-separation principle and Newton-like scheme for coherent quantum LQG control. 18th IFAC World Congress, Milan, Italy, 28 August-2 September, 2011, pp. 4721-4727. (arXiv:1010.3125v2 [quant-ph])

\bibitem{VP11b}
Vladimirov I.G, Petersen I R. A dynamic programming approach to finite-horizon coherent quantum LQG control. Proc. Australian Control Conference, Melbourne, 10-11 November, 2011, pp. 357-362. (arXiv:1105.1574v1 [quant-ph]).


\bibitem{YD84}
Yurke B, Denker J S. Quantum network theory. Phys. Rev. A, 1984, 29:1419-143


\bibitem{ZJ11}
Zhang G, James M R. Direct and indirect couplings in coherent feedback control of linear quantum systems. IEEE Trans. Automat. Contr., 2011, 56(7):1535-1550

\bibitem{ZJ11b}
Zhang G, James M R. On the response of linear quantum stochastic systems to single-photon inputs and pulse-shaping of photon wave packets. Proceedings of the 2011 Australian Control Conference, pp. 55-60, Melbourne, Australia, November 10-11, 2011


\bibitem{ZWL11}
Zhang J, Wu R B, Liu Y, Li C W, Tarn T J. Quantum coherent nonlinear feedbacks with applications to quantum optics on chip. arXiv:1102.2199v2 [quant-ph], 2011

\end{thebibliography}
\end{document}